\documentclass{jfm}
\usepackage{graphicx}
\usepackage{epstopdf, epsfig}
\usepackage{natbib}
\usepackage{hyperref}
\hypersetup{
    colorlinks = true,
    urlcolor   = blue,
    citecolor  = black,
}

\newcommand{\RomanNumeralCaps}[1]
\linenumbers

\usepackage{color}

\usepackage{upgreek}

\shorttitle{The fluid dynamics of collective vortex structures of plant-animal worms}
\shortauthor{G. T. Fortune et. al.}

\title{The fluid dynamics of collective vortex structures of plant-animal worms}

\author{George T. Fortune\aff{1},
 Alan Worley\aff{2},
 Ana B. Sendova-Franks\aff{2},\\
 Nigel R. Franks\aff{2}, Kyriacos C. Leptos\aff{1}, Eric Lauga\aff{1},\\
 \and Raymond E. Goldstein\aff{1}}

\affiliation{
	\aff{1}Department of Applied Mathematics and Theoretical Physics, Centre for Mathematical Sciences, Wilberforce Road, University of Cambridge, Cambridge CB3 0WA, UK
    \aff{2}School of Biological Sciences, University of Bristol, 24 Tyndall Avenue, Bristol BS8 1TQ, UK 
}
\begin{document}

\maketitle

\begin{abstract}
	
Circular milling, a stunning manifestation of collective motion, is found across the natural world, from fish shoals to 
army ants. It has been observed recently that the plant-animal worm \textit{Symsagittifera roscoffensis} exhibits 
circular milling behaviour, both in shallow pools at the beach and in Petri dishes in the laboratory. 
Here we investigate this phenomenon, through experiment and theory, from a fluid dynamical viewpoint, focusing 
on the effect that an established circular mill has on the surrounding fluid.  Unlike systems 
such as confined bacterial suspensions and collections of molecular motors and filaments that exhibit spontaneous
circulatory behaviour, and which are modelled as force dipoles, the front-back symmetry of individual worms 
precludes a stresslet contribution. Instead, singularities such as source dipoles and Stokes quadrupoles are expected to
dominate.  A series of models is analyzed to understand the contributions of these singularities to the azimuthal 
flow fields generated by a mill, in light of the particular boundary conditions that hold for flow in a Petri dish.
A model that treats a circular mill as a rigid rotating disc that generates a Stokes flow 
is shown to capture basic experimental results well, and gives insights into the emergence and stability of multiple mill systems. 
\end{abstract}

\section{Introduction}
          

From the flocking of birds to the schooling of fish, collective motion, global group dynamics resulting 
from the interactions of many individuals, occurs all across the natural world. A visually striking example of 
this is collective vortex behaviour \textemdash the spontaneous motion of large numbers of organisms moving 
in periodic orbits about a common centre. Studied for over a century since Jean-Henri \citet{Fabre99} 
first reported the spontaneous formation of continuous loops in columns of pine processionary caterpillars, 
circular milling has been observed in many species, including army ants 
\citep{Couzin03}, the bacterium \textit{Bacillus subtilis} \citep{Cisneros07,Wioland} and fish \citep{Calovi14}.

\begin{figure}
	\centering
	\includegraphics[trim={0 0cm 0 0cm}, clip, width=\textwidth]{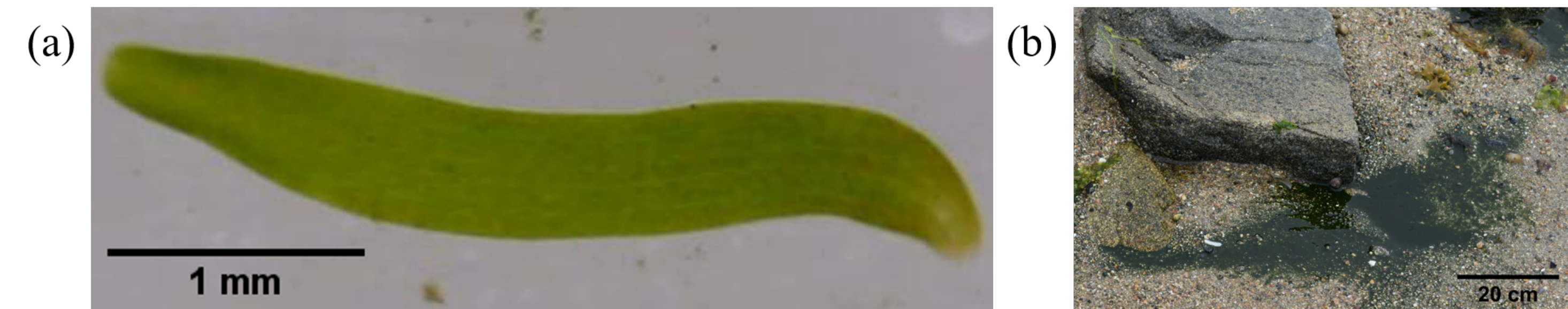}
	\caption{The plant-animal worm \textit{Symsagittifera roscoffensis}. (a) Magnified view of adult. 
	(b) \textit{S. roscoffensis} \textit{in situ} on the beach.}
	\label{fig:1}
\end{figure}

It has been discovered recently that the marine acoel worm \textit{Symsagittifera roscoffensis} \citep{Bourlat09} 
forms circular mills, both naturally in rivulets on intertidal sand \citep{SendovaFranks18}, and in a shallow layer 
of sea water in a Petri dish \citep{Franks16}. \textit{S. roscoffensis}, more commonly known as the `plant-animal 
worm' (figure \ref{fig:1}(a), \cite{Keeble10}), engages in a photosymbiotic relationship \citep{Bailly14} with the 
marine alga \textit{Tetraselmis convolutae} \citep{Norris80}. The photosynthetic activity of the algae 
\textit{in hospite} under the worm epidermis provides the nutrients required to sustain the host. The worms reside 
on the upper part of the foreshore (regions which are typically underwater for around two hours 
before and after high tide) of Atlantic coast beaches in colonies of many millions
(figure \ref{fig:1} (b)). It is hypothesised that this circular milling allows worms to 
self-organise into dense biofilms that, covered by a mucus layer, optimise the absorption of light by the algae 
for photosynthesis \citep{Franks16}.

Here, motivated by prior experimental work and in light of new results we consider a range of issues surrounding the
fluid dynamical description of mills, with particular attention to the fluid velocity field that is 
generated by a circular mill and the 
effect that this flow has on the mill itself. In \S\ref{millingexperiments}, we describe the parameters of field experiments on 
mills performed on the Isle of Guernsey and outline the range of questions they pose.  Systems such as bacterial suspensions \citep{Woodhouse,Wioland},
collections of sperm cells near surfaces \citep{Riedel} 
and assemblies of molecular motors and biofilaments \citep{Sumino} can spontaneously form vortex-like
patterns superficially similar to worm mills. However, their theoretical descriptions \citep{SaintillanShelley} 
treat the constituents as force dipoles 
(stresslets). The front-back symmetry of ciliated worms would suggest that a stresslet contribution is
small if not entirely absent and thus higher-order singularities such as source dipoles ought to appear.
Such is the case with the spherical alga {\it Volvox} whose flow field has
been measured in great detail \citep{Direct}.  As there has been little if any work on the collective behaviour of
suspensions of singularities beyond stresslets, \S\ref{wormbackground} provides background theoretical 
considerations on this problem. First, a detailed examination of the 
relationship between the cilia-generated flow over the surface
of an individual worm and the far-field flow behaviour is given within a prolate squirmer model, with which 
we confirm the absence of
a stresslet contribution for a suitably symmetric surface slip velocity and show that the far field is 
dominated by the source dipole and force quadrupole contributions.  Insight into those
singularity components that lead to azimuthal flow around a mill composed of such swimmers is obtained by 
averaging the far-field behaviour over a circular orbit, which is equivalent to considering a ring of swimmers.  
Within the squirmer model we find that it is the 
Stokes quadrupole that gives the leading-order contribution.  A model of a complete mill can be constructed 
from a suitable oblate squirmer, whose far-field behaviour is that of a rotlet dipole.  With the
particular boundary 
conditions that hold in a Petri dish the far field from this singularity decays exponentially with $z$ dependence $\sin{(\pi z/2H)}$. 
Then in $\S$\ref{millingmaths}, we consider a model of a mill as a rigid disc with varying radius $c = c(t)$ rotating in a 
Stokes flow. A lubrication analysis
of a highly-simplified model of a mill is presented as motivation, and to elucidate the effect of the boundary conditions
for this system. Hence, we vertically average 
the governing equations by setting the $z$ dependence explicitly to $\sin{(\pi z/2H)}$, deriving a Brinkman-like 
equation for the vertically averaged velocity flow field. In general, further analytic progress can not 
be made. However in two particular limits, namely when the mill is close to and when the mill is far away 
from the centre of the Petri dish, an analytic solution for the fluid velocity field and hence for the 
force that the flow exerts on the disc can be derived. In $\S$\ref{millcentredrift}, we demonstrate the 
strong agreement between what is predicted by the model and what is observed experimentally. In particular, 
the viscous force on the disc points in the direction perpendicular to the line between the centres of the mill and 
of the Petri dish. Hence the centre of the disc will drift on a circle with centre the middle of the Petri 
dish, precisely as observed experimentally. 

Finally, in $\S$\ref{binarystagnation} we extend the analysis to systems with more than one mill, focusing on the 
simplest binary mill structure. We utilise the knowledge gained from $\S$\ref{millingmaths} to explain from a 
purely fluid dynamical viewpoint a large raft of experimental observations, including where a second mill forms and 
in what direction it rotates, and the conditions for which a second mill will not form. We can also make predictions 
for the stability of the resulting binary circular mill systems.

\section{Experimental Methods} 
\label{millingexperiments}

\begin{figure}
	\centering
	\includegraphics[trim={0 0cm 0 0cm}, clip, width=\textwidth]{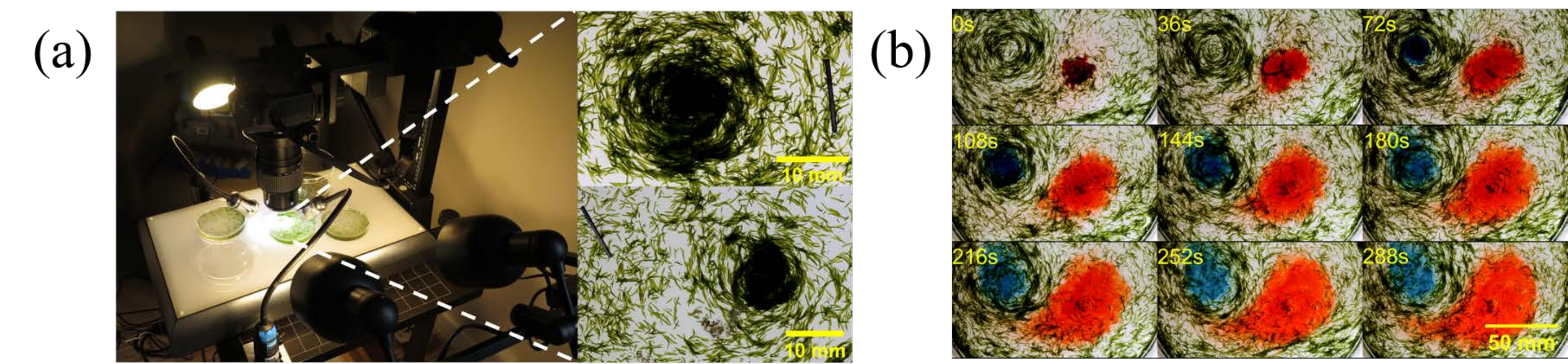}
	\caption{Field experiments. (a) Setup used to film milling behaviour in Guernsey. 
	(b) Montage of still images capturing streaklines produced by the flow.}
	\label{fig:2}
\end{figure}

Here we describe field experiments done during 12-19 June, 2019 in the Peninsula Hotel Guernsey on the Isle of 
Guernsey, a channel 
island near the coast of France.  Worms were collected from a nearby beach 
($49^\circ 29^\prime 45.3^{\prime\prime}$N $2^\circ 33^\prime 14.3^{\prime\prime}$W) 
just prior to the 
experiments, minimising the perturbations in the worms' physiology and behaviour resulting from 
removal from their natural environment.
As shown in figure \ref{fig:2}(a), Petri dishes of diameter $10$ cm were filled with sea water up to a 
depth of $8$ mm. Around ten thousand worms were placed into the dish using $3$ ml plastic Pasteur pipettes. The subsequent 
evolution of the system, including the spontaneous formation of circular mills, was recorded 
at $25$ fps using a Canon Eos 5d Mark II camera equipped with a Canon macro 
lens MP-E $65$ mm f/$2.8$ and a $1.5\times$ magnifier, mounted above the 
dish on a copy stand. The system was illuminated uniformly through by a light box below the Petri dishes 
and LED lights located around them. 

In some experiments, small drops of azo dye were injected into the Petri dish using a plastic Pasteur pipette to act as a 
tracer to track the motion of the fluid. Figure \ref{fig:2}(b) is a montage showing the temporal evolution of a 
red dyed region of fluid, namely streaklines of the flow. As can be seen, the circular mill generates 
a clockwise flow that is in the opposite direction to the anticlockwise direction of rotation of the worms, 
that is,
a backflow generated by the worms pushing themselves through the fluid.

An instantaneous image of a mill shows that its edge is not well defined.  In order to overcome this, every hundred 
frames (i.e. four seconds of footage) were averaged together to create a coarser time lapse video. This averaging sharpens 
the mill edges considerably since this process differentiates between worms entering or leaving the mill and worms 
actually in the mill. Then, the location and radius of the circular mill in each frame were extracted manually, 
utilising a GUI interface in MATLAB to semi-automate this process. 

Appendix A collects relevant information on the many experiments carried out
in the field.  Selected videos can be found in the Supplementary Material.

Circular milling in this system has not previously been studied using the kinds of methods now common 
in the study of active matter \citep{activematter}.  There are open experimental questions at various levels of organisation in
this setup that mirror those that have been successfully answered for bacterial, algal and other microswimmer systems, 
including measurements of flow fields around individual swimmers, pairwise interactions between them, the temporal dynamics of 
mill formation from individuals, the flow fields around the mills and the dynamics of the mills themselves within
their confining containers.  Here our focus experimentally is on the latter; the drift of a mill centre within a Petri
dish and the formation of binary mill systems.

\section{From Individual Worms to Mills} \label{wormbackground}
We begin with fluid dynamical considerations at the level of individual worms to derive key results 
that will then be utilised in \S\ref{millingmaths} to motivate a mathematical model for a circular mill. 
Working in modified prolate spheroidal coordinates, we find that the leading order fluid 
velocity in the far field produced by the locomotion of a single worm can be expressed in terms of 
fundamental Stokes flow point singularities as the superposition of a source dipole and a Stokes 
quadrupole. Among other implications, this result shows that the proper Reynolds number for worm 
locomotion is given by the swimming speed $U$, length $\ell$ and diameter $\rho$ as 
$U(\ell\rho^2)^{1/3}/2\nu \approx 0.36$, so worms swim in the laminar regime. We then consider 
two possible models for a circular mill. Picturing a mill as the superposition of many rings of worms, we 
find that the resulting net flow is azimuthal, that is, not in the vertical $z$ direction. Alternately, considering a mill as an oblate squirmer with axisymmetric swirl, we find that away from the mill, the 
forcing can be expressed as a rotlet dipole and thus the flow has $z$ dependence of the form $\sin(\pi z/2H)$.  

\subsection{Locomotion of an individual worm} \label{singlewormswimming}

The individuals of \textit{S. roscoffensis} studied in the present experiments have a broad distribution of 
sizes; their length $\ell$ ranges from $\approx 1.5-6$ mm, with mean $\bar\ell=3\,$ mm, and diameters $\rho$ 
falling in the range $\approx 0.2-0.6$ mm, with mean $\bar\rho\approx 0.35$ mm. Worm locomotion arises from 
the collective action of carpets of cilia over the entire body surface, each $\approx 20\, \upmu$m long, beating 
at $\approx 50$ Hz. Muscles within the organism allow it to bend, and thereby alter its swimming direction
(\cite{Bailly14}). In unbounded fluid, their average swimming speed of individuals is $U\approx 2$ mm/s. 

To model the fluid velocity field produced by a worm, we follow \cite{Pohnl20} 
and consider a spheroidal, rigid and impermeable squirmer \citep{Lighthill} with semi-minor axis 
$b_x$ and semi-major axis $b_z$ swimming at speed $\boldsymbol{U} = U\boldsymbol{e_z}$ so the 
$z$-axis lies along the major axis. The squirmer moves through prescribing a tangential 
slip velocity $\boldsymbol{u_s}$ at its surface $\Sigma$. Neglecting inertia, the fluid flow in the swimmer frame $\boldsymbol{u}$ satisfies
\begin{equation}
    \mu \nabla^2 \boldsymbol{u} = \nabla p,
\end{equation}
with boundary conditions
\begin{equation}
    \boldsymbol{u} \Big{|}_{\Sigma} = \boldsymbol{u_s}, \quad \boldsymbol{u}\left(|r| \rightarrow \infty \right) \rightarrow - \boldsymbol{U}, \label{eq:swimmingwormb1}
\end{equation}
together with the force-free condition 
\begin{equation}
    \int_{\Sigma} \boldsymbol{\sigma \cdot n} \, d\Sigma = 0, \label{eq:swimmingwormbc2}
\end{equation}
where $\boldsymbol{\sigma}$ is the stress tensor. We now switch to the modified prolate spheroidal coordinates $(\tau, \xi, \varphi)$, utilising the transformations
\begin{subequations}
\begin{equation}
    \tau = \frac{1}{2c}\left( \sqrt{x^2 + y^2 + (z + c)^2} + \sqrt{x^2 + y^2 + (z - c)^2} \right),
\end{equation}
\begin{equation}
    \xi = \frac{1}{2c}\left( \sqrt{x^2 + y^2 + (z+c)^2} - \sqrt{x^2 + y^2 + (z - c)^2} \right),
\end{equation}
\begin{equation}
    \varphi = \arctan\left(\frac{y}{x}\right),
\end{equation} \label{eq:prolatespheroidalcoordinates}%
\end{subequations}
with $c = \sqrt{b_z^2 - b_x^2}$ and the squirmer boundary is mapped to the surface $\tau = \tau_0 = b_z/c =$ constant. In this coordinate system, assuming an axisymmetric flow $\boldsymbol{u} = u_{\tau}\boldsymbol{e_{\tau}} + u_{\xi}\boldsymbol{e_{\xi}}$ and axisymmetric tangential slip velocity $\boldsymbol{u_s} = u_s\boldsymbol{e_{\xi}}$, the Stokes streamfunction $\psi$ satisfies
\begin{subequations}
\begin{equation}
    u_{\tau} = \frac{1}{h_{\xi}h_{\varphi}}\frac{\partial \psi}{\partial \xi} = \frac{1}{c^2 \sqrt{(\tau^2 - \xi^2)(\tau^2 - 1)}}\frac{\partial \psi}{\partial \xi},
\end{equation}
\begin{equation}
  u_{\xi} = -\frac{1}{h_{\tau}h_{\varphi}}\frac{\partial \psi}{\partial \tau} = - \frac{1}{c^2 \sqrt{(1 - \xi^2)(\tau^2 - \xi^2)}}\frac{\partial \psi}{\partial \tau}.  
\end{equation}
\end{subequations}
Taking from \cite{Dassios94} the general separable solution for the stream function in prolate spheroidal coordinates and applying the boundary conditions given in (\ref{eq:swimmingwormb1}) and (\ref{eq:swimmingwormbc2}), as in \cite{Pohnl20}, we obtain
\begin{equation}
\psi(\tau,\xi) = \sum_{n = 2}^{\infty} g_n(\tau)G_n(\xi),
\end{equation}
where the $g_n$ satisfy
\begin{subequations}
\begin{equation}
    g_2(\tau) = C_4H_4(\tau) + D_2H_2(\tau) - 2c^2U G_2(\tau),
\end{equation}
\begin{equation}
    g_3(\tau) = -\frac{C_3}{90}G_0(\tau) + C_5H_5(\tau) + D_3H_3(\tau),
\end{equation}
\begin{equation}
    g_{n \geq 4}(\tau) = C_{n + 2}H_{n + 2}(\tau) + C_nH_{n - 2}(\tau) + D_nH_n(\tau),
\end{equation}
\end{subequations}
where $G_n$ and $H_n$ are Gegenbauer functions of the first and second kind respectively and 
$P^1_n = -\sqrt{1 - x^2}dP_n/dx$ are the associated Legendre polynomials.  The integration 
constants $C_n$ and $D_n$ are set by the boundary conditions
\begin{equation}
    g_n(\tau_0) = 0 \, \mbox{ and } \, \frac{d g_n}{d\tau}\Big{|}_{\tau = \tau_0} = \tau_0 c^2 n(n-1)B_{n-1} \, \mbox{ for } \, n = 2,3 \cdots \label{eq:gnbc}
\end{equation}
Here, the $\{B_n\}_{n \geq 1}$ are the coefficients in the series expansion of $u_s = \tau_0 \sum_{n \geq 1} B_n V_n(\xi)$ using the set of functions $\{ V_n = (\tau_0^2 - \xi^2)^{-1/2}P_n^1(\xi) \}_{n \geq 1}$, which is a basis over the space of $L^2$ functions satisfying $f(\xi = \pm 1) = 0$ together with the inner product $<>_w = \int^1_{-1} \, w \, d\xi$ and the weight function $w = \tau_0^2 - \xi^2$. Furthermore, the swimming speed $U$ can be expressed in terms of the $\{B_n\}_{n \geq 1}$ using
\begin{equation}
    U = -\frac{\tau_0}{2}\int^1_{-1} \frac{\sqrt{1 - \xi^2}}{\sqrt{\tau_0^2 - \xi^2}} v_s(\xi) \, d\xi = \frac{\tau_0^2}{2} \sum_{\mbox{$n$ odd}} B_nU_n \mbox{ where } U_n = \int^1_{-1} \frac{P_1^1 P_n^1}{\tau_0^2 - x^2}\, dx, \label{eq:Uexpression}
\end{equation}
namely only odd enumerated modes contribute to the squirmer's swimming velocity. Hence, from now on we only consider the case when the forcing is only a linear combination of the odd modes i.e. $B_{2n+2} = 0 \rightarrow g_{2n+1} = 0 \rightarrow C_{2n+1} = D_{2n+1} = 0 \, \forall n \in \mathbb{Z}^{+}$. When the prescribed forcing arises purely from the first mode, i.e.  $u_s = \tau_0 B_1V_1$, $C_{2n \colon n \geq 1}$ and $D_{2n \colon n \geq 1}$ simplify to become 
\begin{equation}
    D_2 = -2B_1c^2\tau_0(\tau_0^2 - 1) \, \mbox{ and } \,  C_{2n \colon n \geq 1} = D_{2n \colon n \geq 2} = 0.
\end{equation}
When the forcing arises from a higher order mode, i.e. $u_s = \tau_0 B_{2n+1}V_{2n+1}$ where $n \geq 1$, $D_2$ and $C_4$ simplify to become
\begin{eqnarray}
\left( 
    \begin{array}{ll}
     C_4 \\
     D_2
     \end{array} \right) &=& \frac{B_{2n+1}c^2U}{H_4(\tau_0)H_2'(\tau_0)-H_2(\tau_0)H_4'(\tau_0)} \nonumber \\
     &\times& \left( 
    \begin{array}{ll}
     1 \\
     \frac{2}{3} + \frac{5\tau_0^4}{4} - \frac{25\tau_0^2}{12} - \frac{5\tau_0}{8}(1 - \tau_0^2)^2\log{\left(\frac{\tau_0+1}{\tau_0-1}\right)}
     \end{array} \right),
\end{eqnarray}
i.e. the only $n$ dependence arises from the $U$. Moving into the lab frame, in the far field ($\tau \gg 1$) the dominant term in the expansion for $\psi$ comes from $H_{2}(\tau) = 1/3\tau + \cdots$, so
\begin{subequations}
\begin{equation}
    \psi = \frac{1}{3\tau} \left( \frac{D_2}{2}\left(1-\xi^2\right) - \frac{C_4}{8}\left(1 - 6\xi^2 + 5\xi^4\right) \right) + \cdots,
\end{equation}
\begin{eqnarray}
    u_{\xi} &=& -\frac{1}{\tau c^2 \sqrt{1 - \xi^2}}\frac{\partial \psi}{\partial r} + \cdots\nonumber \\
    &=& \frac{1}{3\tau^3c^2\sqrt{1 - \xi^2}}\left( \frac{D_2}{2}\left(1-\xi^2\right) - \frac{C_4}{8}\left(1 - 6\xi^2 + 5\xi^4\right) \right) + \cdots,
\end{eqnarray}
\begin{equation}
    u_{\tau} = \frac{1}{\tau^2c^2}\frac{\partial \psi}{\partial \xi} + \cdots = -\frac{\xi}{3\tau^3c^2}\left( D_2 + \frac{C_4}{2}\left(5\xi^2 - 3\right) \right) + \cdots,
\end{equation} \label{eq:farfieldgeneralcase}
\end{subequations}
Converting this back to vector notation yields
\begin{equation}
    \boldsymbol{u} = -\frac{1}{2c^2}\left[ \left(D_2 + \frac{C_4}{6}\right)\boldsymbol{u_D} + \frac{5C_4}{36}\boldsymbol{u_Q} \right] + \cdots \label{eq:fullfarfield}
\end{equation}
where $\boldsymbol{u_D}$ and $\boldsymbol{u_Q}$, the flows generated by a source dipole and a Stokes quadrupole respectively, satisfy
\begin{subequations}
\begin{eqnarray}
    \boldsymbol{u_D} &=& \left( \frac{c}{r} \right)^3\left( \frac{x_ix_k}{r^2} - \frac{\delta_{ik}}{3} \right), \label{eq:farfielddipole} \\
    \boldsymbol{u_Q} &=& \frac{\partial}{\partial x_k^2}\left( \frac{x_ix_k}{r^3} + \frac{\delta_{ik}}{r} \right) \nonumber \\
    &=& 3\left( \frac{c}{r} \right)^3\left( \frac{5x_ix_k^2}{r^2} - (2 + \delta_{ik})\frac{x_ix_k}{r^2} - \left( \frac{x_ix_k}{r^2} - \frac{\delta_{ik}}{3} \right) \right). \label{eq:farfieldquadrupole}
\end{eqnarray}
\end{subequations}
Thus, the far field fluid velocity field decays like $1/r^3$, consisting of a combination of a source 
dipole and a Stokes quadrupole. The far field fluid velocity generated by a higher order than one odd mode squirmer 
contains both source dipole and quadrupole components with the quadrupole component dominating as $\tau_0 \rightarrow 1$. 

Similarly, the far field fluid velocity generated by a mode one squirmer is purely a source dipole.  
Using \cite{Lauga20}, this is the same as an efficient spherical squirmer (forcing only arising from the first mode) 
with effective radius $\tilde{a} = c^3\tau_0(\tau_0^2 -1)$. When $\tau_0 \rightarrow \infty$, namely the spherical limit, 
as expected $\tilde{a} \rightarrow b_x = b_z = a$, the radius of the sphere. When $\tau_0 \rightarrow 1$, 
the elongated limit, $\tilde{a} \rightarrow ( b_z b_x^2 )^{1/3}$. Thus, at the scale of an individual worm, 
the Reynolds number $U \tilde{a}/\nu$ in water ($\nu=1\,$ mm$^2$/s) is $\approx\! 0.36$ where 
$\tilde{a} = ( b_z b_x^2 )^{1/3} = (\bar\ell \bar\rho^2)^{1/3}/2$ is the correct length scale for locomotion of an 
individual worm. Inertial effects are modest and individual worms swim in the laminar regime.

\subsection{Ring of Spheriodal Squirmers} \label{ringofsquirmers}
Given the results in above, it is natural to ask which singularities associated with individual worms contribute
to the azimuthal flow around a mill.  This can be investigated by averaging over the contributions from a swimmer
in a circular orbit, as has been done in the stresslet case \citep{Michelin10}.  Hence, 
consider a spheroidal squirmer swimming clockwise horizontally in a circle of radius $c$, instantaneously located at 
the point $P = (c \cos{\theta}, c \sin{\theta},0)$ and orientated in the direction 
$\boldsymbol{p} = (\sin{\theta}, -\cos{\theta},0)$, utilising a Cartesian coordinate system $(x,y,z)$ 
with origin at the centre of the circle. 
If each squirmer generates a source dipole $\boldsymbol{u_{D}}$, the fluid velocity  
$\boldsymbol{u_{DX}}(c,\theta)$ at $X = (R,0,0)$ is
\begin{equation}
\boldsymbol{u_{DX}} = \frac{c^3}{3r^5}\left( 
    \begin{array}{ll}
     \left(2R^2-c^2\right)\sin{\theta} - cR\sin{\theta}\cos{\theta} \\
     \left(R^2 + c^2\right)\cos{\theta} - cR(3 - \cos^2{\theta})
     \end{array} \right),
\end{equation}
where $r = (R^2 + c^2 - 2cR\cos{\theta})^{1/2}$. The total velocity  
$\boldsymbol{u_{Dring}}(R)$ at $X$ due to a ring of clockwise swimming worms of radius $c$ with line density $\lambda_{ring}$ is then
\begin{equation}
    \boldsymbol{u_{Dring}}(R) = c \, \lambda_{ring} \int^{\pi}_{-\pi} \boldsymbol{u_{wd}}(c,\theta) 
    \, d\theta = 0. \label{eq:flowfielddipolering}
\end{equation}
Hence, a ring of uniformly distributed source dipole swimmers generates no net flow field 
outside of the ring. By contrast, the flow field $\boldsymbol{u_{Qring}}(R)$ due
to a ring of clockwise swimming worms, each generating 
a Stokes quadrupole, is finite:
\begin{equation}
\boldsymbol{u_{Qring}}(R) = c\lambda_{ring}\left( \frac{c}{R} \right)^3 \Upsilon\left( \frac{c}{R} \right)\boldsymbol{e_y}, \label{eq:flowfieldquadrupolering}
\end{equation}
where
\begin{eqnarray}
    \Upsilon(x) &=& \int^{2\pi}_0 \frac{\cos{\theta}\left( 4-23x^2-x^4 +\cos^2{\theta}\left(17x^2-5\right)\right)}{(1 + x^2 - 2x\cos{\theta})^{7/2}} \, d\theta \nonumber \\
    &-& \int^{2\pi}_0 \frac{x\left( 12 - 13x^2 + \cos^2{\theta}\left( 9x^2-31 \right) + 15\cos^4{\theta} \right)}{(1 + x^2 - 2x\cos{\theta})^{7/2}} \, d\theta.
\end{eqnarray}
This decays in the far field as $1/R^4$.  We conclude that, viewing the mill in terms of its individual
constituents, it is the Stokes quadrupole from individual
swimmers that drives the dominant azimuthal flow.  

\subsection{Oblate Squirmer with Swirl} \label{swirlingsquirmer}

Further insight into the flow field generated by a mill can be obtained by viewing it as a single,
self-propelled object with some distribution of velocity on its surface arising from the many cilia
of the constituent worms.  With a shape like a pancake, it can be modelled as an 
oblate squirmer with axisymmetric swirl.
First, consider a prolate squirmer with aspect ratio $r_e = b_x/b_z$ rotating in the 
$\varphi$ direction with imposed surface flow $u_{\varphi 0}(\xi) \, \boldsymbol{e_{\varphi}}$ in free space. Assuming that the generated fluid flow is purely in the $\varphi$ direction with no $\varphi$ dependence, the $\varphi$ component of the Stokes equations, $\mu (\nabla^2 \boldsymbol{u})_{\varphi} = 0$, becomes
\begin{equation}
(\tau^2 - 1)\frac{\partial^2 u_{\varphi}}{\partial \tau^2} + 2\tau\frac{\partial u_{\varphi}}{\partial \tau} - \frac{u_{\varphi}}{\tau^2 - 1} + (1 - \xi^2)\frac{\partial^2 u_{\varphi}}{\partial \xi^2} - 2\xi\frac{\partial u_{\varphi}}{\partial \xi} - \frac{u_{\varphi}}{1 - \xi^2} = 0. \label{eq:stokesvarphiterm}
\end{equation}
This admits the general separable solution that tends to zero at infinity
\begin{equation}
    u_{\varphi} = \sum_{n = 1}^{\infty} C_{pn} P^1_{n}(\xi) \, Q^1_{n}(\tau), \label{eq:spheriodalseparablesolution}
\end{equation}
where $C_{pn}$ are constants and as before $P^1_n(\xi)$ and $Q^1_n(\tau)$ are associated Legendre 
polynomials. Furthermore, since the squirmer is force and torque free, $C_{p1} = 0$. Decomposing 
$u_{\varphi 0}$ using the basis $\{ V_n(\xi) \}$, i.e. $u_{\varphi 0} = 
\sum_{n = 2}^{\infty} C_{n0} V_n(\xi)$, we find that $C_{pn} = C_{n0}/W_{pn}(\tau_0)$ where 
$\tau_0 = 1/\sqrt{1 - (r_e)^2}$ and $r_e = b_x/b_z \leq 1$ is the aspect ratio of the spheroid. 
Note that in the spherical limit ($r_e \rightarrow 1$, $b_x = b_z = a$) 
(\ref{eq:spheriodalseparablesolution}) simplifies to become
\begin{equation}
    u_{\varphi} = \sum_{n = 1} a \overline{C}_n \frac{a}{r^{n+1}} V_n(\xi), 
\end{equation}
where $\overline{C}_n$ are constants, and we recover the general form for a spherical squirmer with
swirl \citep{Pak14,Pedley16}. 

Returning to the general case, the dominant term in the far field arises from mode $2$,
\begin{equation}
    u_{\varphi} = -\frac{2C_{p2}}{5\tau^3}\xi\sqrt{1-\xi^2} + \cdots \rightarrow \boldsymbol{u} = \frac{2C_{p2}}{5}\left( \frac{c}{r} \right)^3\frac{\boldsymbol{x \times x_k}}{r^2},
\end{equation}
where $x_k$ points in the z direction.  This is a rotlet dipole. 
Now, using \cite{Dassios94}, to compute the velocity fluid for an oblate squirmer with swirl of 
aspect ratio $r_e' >1$, we translate the results from the prolate spheroidal coordinate 
system $(\tau,\xi,\varphi)$ to the oblate spheroidal coordinate system $(\lambda,\xi,\varphi)$ using 
the substitutions
\begin{equation}
    \tau = i\lambda, \quad c = -i\overline{c},
\end{equation}
where $\overline{c} = \sqrt{b_x^2 - b_z^2}$ and $(\lambda,\xi,\varphi)$ can be expressed in terms of the Cartesian coordinates $(x,y,z)$ using
\begin{subequations}
\begin{equation}
    \lambda = \frac{1}{2\overline{c}}\left( \sqrt{x^2 + y^2 + (z - i\overline{c})^2} + \sqrt{x^2 + y^2 + (z + i\overline{c})^2} \right),
\end{equation}
\begin{equation}
    \xi = \frac{1}{-2i\overline{c}}\left( \sqrt{x^2 + y^2 + (z - i\overline{c})^2} - \sqrt{x^2 + y^2 + (z + i\overline{c})^2} \right),
\end{equation}
\begin{equation}
    \varphi = \arctan{\left( \frac{y}{x} \right)}.
\end{equation}
\end{subequations}
Similarly to the prolate case, in the far field the second mode dominates, giving a fluid velocity field 
also in the form of a rotlet dipole,
\begin{equation}
    \boldsymbol{u} = \frac{2C_{ob2}}{5}\left( \frac{\overline{c}}{r} \right)^3\frac{\boldsymbol{x \times x_k}}{r^2},
    \label{rotletdoublet}
\end{equation}

This result is intuitive; in the absence of a net torque on the object there can not be a rotlet 
contribution, so analogously to the case of a single bacterium whose body rotates opposite to that 
of its rotating helical flagellum, the rotlet dipole is the first nonvanishing rotational 
singularity.

We close this section by asking how a free-space singularity of the type in (\ref{rotletdoublet}) 
is modified when placed in a fluid layer with the boundary conditions of a Petri dish.  Here we quote
from a lengthy discussion \citep{Fortune20} of a number of cases that complements earlier work
\citep{Liron76} on singularities bounded by two no-slip walls; the leading order contribution in the 
far field to the fluid velocity from an oblate squirmer with swirl at $z = h$ between a no-slip lower 
surface at $z = 0$ and an upper free surface at $z = H$ is
\begin{equation}
    \boldsymbol{u_p} = \frac{2\pi^2 C_{ob2}}{15}\left( \frac{\overline{c}}{H} \right)^3\frac{\boldsymbol{x \times \hat{k} }}{\rho}\cos{\left( \frac{\pi h}{2H} \right)}\sin{\left( \frac{\pi z}{2H} \right)}K_1{\left( \frac{\rho \pi}{2H} \right)}, \label{eq:petridishrotletdipole}
\end{equation}
where $\rho^2 = x^2 + y^2$ and $K_1$ is the Bessel function.  The flow decays exponentially away from the squirmer with decay length $2H/\pi$.

\section{Mathematical Model for a Circular Mill} \label{millingmaths}
\subsection{Background} \label{eq:millingbackground}
We now proceed to develop a model the collective vortex structures observed experimentally in \S\ref{millingexperiments}. 
A laboratory mill of the kind studied here typically has a radius $c$ in the range $5-20$ mm and rotates roughly as a rigid 
body with period $T = 2\pi c/U$ in the range $15-60$ s and angular frequency $\omega = U/c$ in the range $0.4-0.1$ s$^{-1}$ 
whereas in \S\ref{singlewormswimming} the average worm swimming speed $U \approx 2$ mm/s. Almost all the worms swim 
just above the bottom of the Petri dish in a layer typically only one worm thick, with even in the densest regions of the mill 
at most two or three worms on top of each other. 

We observe minimal variation in the height $H$ of the water in the Petri dish. Furthermore, by tracking dye
streaklines we also observe minimal fluid flow in the vertical ($z$) direction. These observations are consistent with 
the considerations in \S\ref{wormbackground}; we can picture a circular mill as a superposition of many rings of worms, 
each of which lies in a horizontal plane. Combining (\ref{eq:fullfarfield}), (\ref{eq:flowfielddipolering}) 
and (\ref{eq:flowfieldquadrupolering}), the far field flow field for each ring is azimuthal, not in the 
vertical direction, and thus the net flow for the circular mill is horizontal as well.

As a final reduced model, we make the further simplification of considering a mill as a rotating disc with a defined centre $b(t)$, radius $c(t)$ and height $d(t)$, 
quantities that are allowed to vary as a function of time. To maximise swimming efficiency, isolated worms away from the mill 
will propel themselves mostly from the first order mode $V_1$ given in \S\ref{singlewormswimming}. Since this fluid velocity 
field decays rapidly away from a worm and using \S\ref{ringofsquirmers} is zero across a full orbit, we neglect the fluid 
flow generated by worms away from the mill. Since a mill contains a high density of worms together with the interstitial 
viscoelastic mucus, we assume that the disc is rigid. Finally, since the locomotion of the worms generates a fluid backflow 
in the opposite direction to their motion, the disc is assumed to rotate in the opposite angular direction to that of the worms.

A common mathematical tool for solving problems in a Stokes flow is to express the forcing as the sum of a finite set of 
fundamental Stokes flow point singularities \citep{Jeong92,Crowdy10}. Given in \S\ref{swirlingsquirmer}, by considering 
the circular mill as an rigid oblate squirmer with swirl, the dominant contribution from the forcing can be approximated as 
a rotlet dipole. However from (\ref{eq:petridishrotletdipole}), the leading order contribution in the far field for a 
rotlet dipole trapped between a lower rigid no-slip boundary and an upper free surface which deforms minimally has 
$z$ dependence of the form $\sin{\left(\pi z/2H\right)}$. Hence, we then vertically average the governing equations by 
setting the $z$ dependence of $u$ to be precisely $\sin{\left(\pi z/2H\right)}$ i.e. we employ the factorisation
\begin{equation}
    \boldsymbol{u} = \frac{\pi}{2}\sin{\left(\frac{\pi z}{2H}\right)} \boldsymbol{U}(r),
\end{equation}
where $\boldsymbol{U} = \boldsymbol{U(r)}$ is independent of $z$ and the factor $\pi/2$ is for convenience.

Thus, a suitable rotational Reynolds number on the scale of a mill is $Re \sim UX/\nu \approx 10$ where 
$X = 2H/\pi$.  Moreover, the dominant velocities are azimuthal, with gradients in the radial direction. 
This suggests that the fluid dynamics of milling is certainly in the laminar regime, and 
the neglect of inertial terms is justified.

\begin{figure}
	\centering
	\includegraphics[trim={0 0cm 0 0cm}, clip, width=\textwidth]{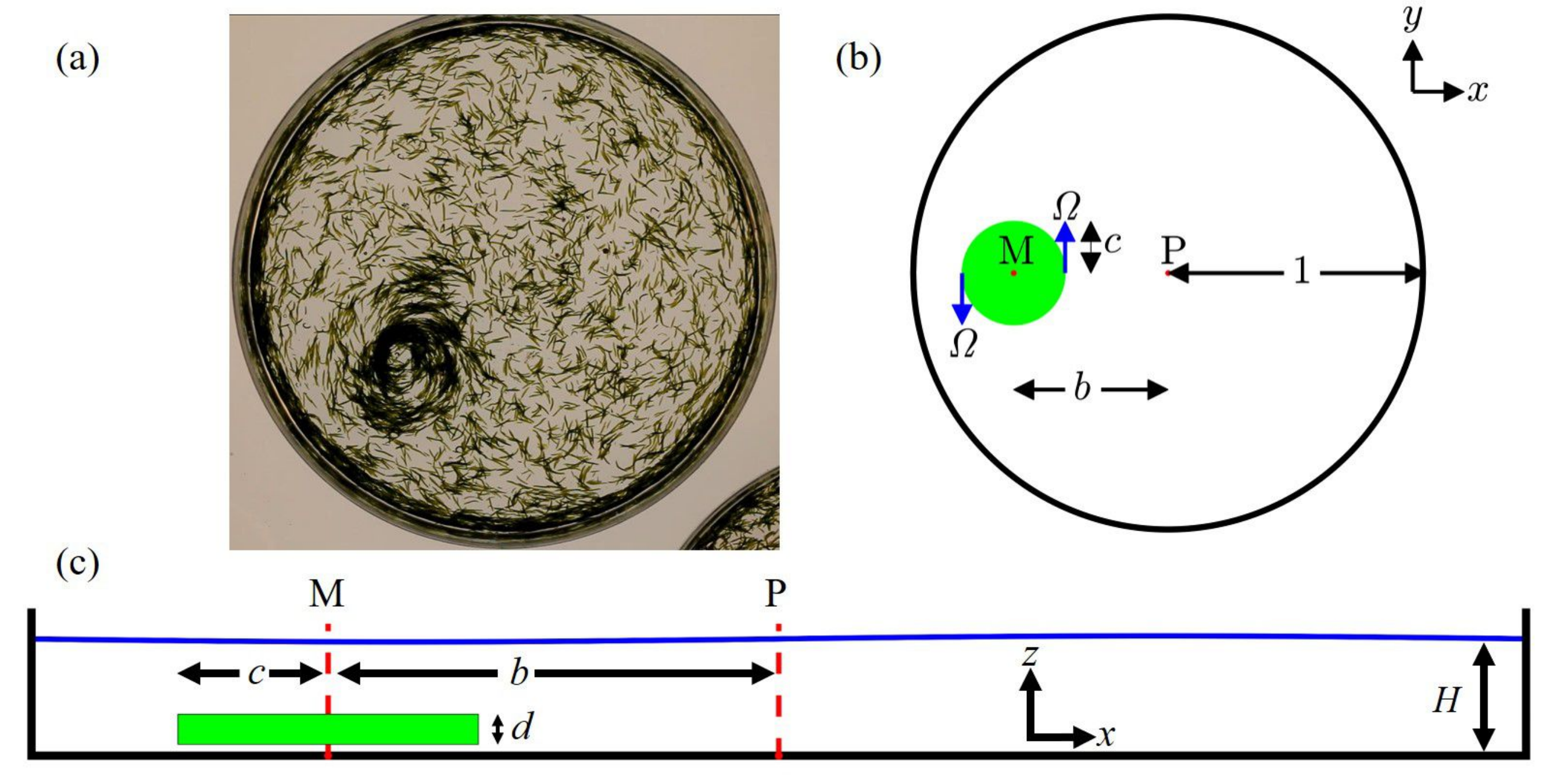}
	\caption{System containing a single mill: (a) Experimental view. (b) + (c) Plan and Front view for the corresponding schematic showing a disc, rotating with angular velocity $\Omega$, which has radius $c$, thickness $d$ and centre M a distance $b$ away from the centre P of a circular Petri dish of unit radius.}
	\label{fig:3}
\end{figure}

\subsection{Defining Notation}

As shown in Figure \ref{fig:3} and in Supplementary Video 1, we define a coordinate basis $(x,y,z)$ with origin 
at the centre of the Petri dish P, where the bottom of the dish is at $z = 0$ and the free surface is at $z = H$, a constant. 
We model an established circular mill, rotating a distance $d_0$ above the bottom of a circular Petri dish with angular 
velocity $-\Omega$, as a rigid disc of radius $c$ and height $d$ with imposed angular velocity $\Omega$, generating a flow 
in a cylindrical domain with cross sectional radius 1 where $d_0 \ll d \ll c \, , \, b \, , \, H$. Let the centre of the disc M have instantaneous position $(-b,0,d_0 + d/2)$.

\subsection{Lubrication Picture}
Insight into the mill dynamics comes first from an extremely simplified calculation within lubrication theory in which
the disc (mill) has a prescribed azimuthal slip velocity on its bottom surface and a simple no-slip condition on its top 
surface, as if only the bottom of worms have beating cilia.   This artiface allows the boundary conditions at $z=0$ and $z=H$
to be satisfied easily. For the thin film of fluid between the bottom of the mill and the bottom of the dish, 
namely $\{ (x,y,z) \colon 0 \leq z \leq d_0;, x^2 + y^2 \leq c^2 \}$, let the imposed slip velocity at $z = d_0$ 
be $\boldsymbol{u_\mathrm{slip}} = u_\mathrm{slip}(r)\boldsymbol{\hat{e}_{\theta}}$. In the absence of any pressure gradients, the general
results of lubrication theory dictate a linear velocity profile for the flow $\boldsymbol{u_b}$ in the film,
\begin{equation}
   \boldsymbol{u_b}(r,z) = \frac{z}{d_0}\left( u_\mathrm{slip} + \Omega r \right)\boldsymbol{\hat{e}_{\theta}},
\end{equation}
where $\Omega$ is the as yet unknown angular velocity of the disc. The flow in the region above the mill is 
simply $u(r,z) = \Omega r$, independent of $z$; it rotates with the disc as
a rigid body.
The torque $T_b$ on the underside of the mill is
\begin{equation}
    T_b = \frac{2\pi\mu}{d_0} \int^c_0 r^2 dr \, (u_\mathrm{slip} + \Omega r), 
 \end{equation}
while there are no torques from the flow above because of its $z$-independence. 
Since the mill as a whole is torque free, $T_b = 0$ and we deduce
\begin{equation}
    \Omega = - \frac{4}{c^4} \int^c_0 r^2 dr \, u_\mathrm{slip}(r).
\end{equation}
If $u_\mathrm{slip}$ has solid body like character, $u_\mathrm{slip} = u_0 r/c$, then $\Omega = -u_0/c$ and $u_b = 0$.  
This ``stealthy" mill generates at leading order no net flow in the gap between the mill and the bottom of the dish and it
is analogous to the stealthy spherical squirmer with swirl that generates no external flow \citep{Lauga20}.  
Any slip velocity other than the solid body form will generate flow in the layer, and one notes generically 
that it is in the opposite direction to the slip velocity.  This is consistent with the phenomenology shown 
in figure \ref{fig:2} involving the
backwards advection of dye injected near a mill. Hence, from now on, we will assume that the mill effectively imposes a constant velocity boundary condition at the edge of the mill, $(x+b)^2 + y^2 = c^2$.
\subsection{Full Governing Equations}

Assuming Stokes flow with fluid velocity $\boldsymbol{u} = (u_x, \, u_y, \, w = 0)$, pressure $p = p(x,y,z)$ and viscosity $\mu$, 
the governing equations are
\begin{equation}
\mu \nabla^2 \boldsymbol{u} = \boldsymbol{\nabla}p \, , \, \boldsymbol{\nabla \cdot u} = 0. \label{eq:stokesflow}
\end{equation} 
Employing no slip boundary conditions (\cite{Batchelor67}) at both the outer edge and the bottom of the Petri dish yield
\begin{equation}
\boldsymbol{u} = 0 \mbox{ at } z = 0 \, , \, \boldsymbol{u} = 0 \mbox{ at } r = \sqrt{x^2 + y^2} =1.
\end{equation}
On the surface of the fluid, $z = H$, the dynamic boundary condition is
\begin{equation}
(\boldsymbol{\sigma_f - \sigma_a}) \boldsymbol{\cdot \hat{z}} = 0, \label{eq:upperdynamicbc}
\end{equation}
where $\boldsymbol{\sigma_f}$ and $\boldsymbol{\sigma_a} = -p_0\boldsymbol{I}$ are stress tensors for the fluid 
and the air respectively. Finally, the boundary conditions on the surface of the mill become
\begin{subequations}
\begin{equation}
    \boldsymbol{u \cdot e_t} = \Omega \tilde{c} \mbox{ on } \Gamma = \{ (x,y,z) \, \colon \, 
    \tilde{c}^2 = (x + b)^2 + y^2 \leq c^2, \, z = d_0 \}, \\
\end{equation}
\begin{equation}
    \boldsymbol{u \cdot e_t} = \Omega c \mbox{ on } \Gamma = \{ (x,y,z) \, \colon \, (x + b)^2 + y^2 = c^2, \, d_0 \leq z \leq d_0 + d \}, \\
\end{equation}
\begin{equation}
    \boldsymbol{u \cdot e_t} = \Omega \tilde{c} \mbox{ on } \Gamma = \{ (x,y,z) \, \colon \, \tilde{c}^2 
    = (x + b)^2 + y^2 \leq c^2, \, z = d_0 + d \},
\end{equation}
\end{subequations}
where the tangent and normal vectors $\boldsymbol{e_t}$ and $\boldsymbol{e_n}$ satisfy
\begin{subequations}
\begin{equation}
   \boldsymbol{e_n} = \frac{1}{\left( y^2 + (x+b)^2 \right)^{1/2}}\left(\left(x+b\right)\boldsymbol{e_x} 
   + y\boldsymbol{e_y}\right), \label{eq:en}
\end{equation}
\begin{equation}
    \boldsymbol{e_t} = \frac{1}{\left( y^2 + (x+b)^2 \right)^{1/2}}\left((-y\boldsymbol{e_x} 
    + \left(x+b\right)\boldsymbol{e_y}\right). \label{eq:et}
\end{equation}
\end{subequations}

\subsection{Vertically-averaged Governing Equations}
Defining $\boldsymbol{r}$ as the in-plane coordinates $(x,y)$, as set out in \S\ref{eq:millingbackground}, we employ the factorisation
\begin{equation}
    \boldsymbol{u} = \frac{\pi}{2}\sin{\left( \frac{\pi z}{2H} \right)}\boldsymbol{U}(\boldsymbol{r}) 
    \longrightarrow \nabla^2 \boldsymbol{u} =  \frac{\pi}{2}\sin{\left( \frac{\pi z}{2H} \right)}\left(\nabla^2 - \kappa^2 \right) \boldsymbol{U},
\end{equation}
where $\kappa=\pi/2H$ plays the role of the inverse Debye screening length in screened electrostatics, 
and the $z$-dependent prefactor guarantees both the lower no slip and the upper stress free vertical boundary conditions. 
Vertically averaging, i.e considering $H^{-1}\int^{H}_{0}\cdots\, dz$, gives the Brinkman-like equation
\begin{equation}
\mu \left( \nabla^2 \boldsymbol - \kappa^2 \right)\boldsymbol{U} = \boldsymbol{\nabla}p \, , 
\ \ \  \boldsymbol{\nabla \cdot U} = 0, \label{eq:verticalaveragedstokesflow}
\end{equation}
$\boldsymbol{U} = U_x \boldsymbol{e_x} + U_y \boldsymbol{e_y} = U_n \boldsymbol{e_n} + U_t \boldsymbol{e_t}$ 
has a corresponding Stokes streamfunction $\varphi$ satisfying
\begin{equation}
    \left \{ U_x = \frac{\partial \varphi}{\partial y}, U_y = - \frac{\partial \varphi}{\partial x} \right \} 
    \longrightarrow \nabla^4 \varphi = \kappa^{2} \nabla^2 \varphi, \label{eq:streamfunctioneq}
\end{equation}
together with boundary conditions
\begin{subequations}
\begin{equation}
u_n = 0 \mbox{ at } r = \sqrt{x^2 + y^2} =1,
\end{equation}
\begin{equation}
u_t = 0 \mbox{ at } r = \sqrt{x^2 + y^2} =1,
\end{equation}
\begin{equation}
    u_n = 0 \mbox{ on } \Gamma = \{ (x,y) \, \colon \, (x+b)^2 + y^2 = c^2 \}, \label{eq:gammaun}
\end{equation}
\begin{equation}
    u_t = \Omega c \mbox{ on } \Gamma = \{ (x,y) \, \colon \, (x+b)^2 + y^2 = c^2 \}, \label{eq:gammaut} 
\end{equation}
\end{subequations}
i.e. no-slip is imposed at the edge of the Petri dish while a constant azimuthal velocity boundary condition is imposed at $\{ (x,y) \, \colon \, (x+b)^2 + y^2 = c^2 \}$. Using separation of variables, the general series solution in polar coordinates to  \ref{eq:streamfunctioneq} is
\begin{eqnarray}
    \varphi &=& A_0 + B_0 \log{r} + C_0 I_0(\kappa r) + D_0 K_0(\kappa r) \nonumber \\
    &+& \sum_{n = 1}^{\infty} \cos{n \theta}\left( A_n r^n + B_n r^{-n} + C_n I_n(\kappa r) + D_n K_n(\kappa r) \right) \nonumber \\
    &+& \sum_{n = 1}^{\infty} \sin{n \theta}\left( \tilde{A}_n r^n + \tilde{B}_n r^{-n} + \tilde{C}_n I_n(\kappa r) 
    + \tilde{D}_n K_n(\kappa r) \right), \label{eq:separableform}
\end{eqnarray}
where $\{ I_n(r) \, , \, K_n(r) \}$ are solutions of the first and second kind respectively for the modified Bessel 
equation $f'' + f'/r - f(1 + n^2/r^2) = 0$. In general, this system does not admit an analytic solution. However, 
significant analytic progress can be made in two particular limits, namely when the mill is close to and when the mill 
is far away from the centre of the Petri dish.

\subsection{Near-field Perturbation Analysis} \label{nearfieldpeturb}

\begin{figure}
	\centering
	\includegraphics[trim={0 0cm 0 0cm}, clip, width=\textwidth]{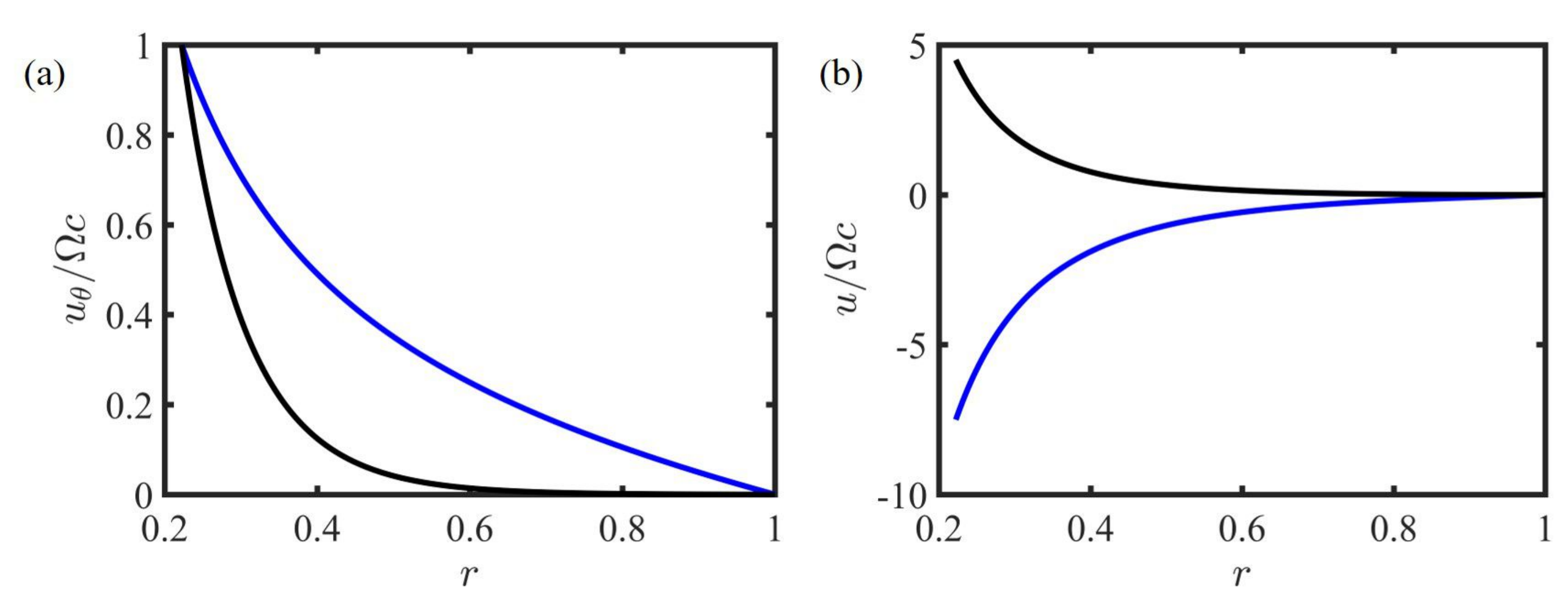}
	\caption{(a) Azimuthal fluid velocity profile, plotted as a function of distance from the Petri dish centre $r$, 
	for both the Brinkman solution (black) and the corresponding Couette solution (blue) when $c = 10/45$ and $h = 8/45$. 
	(b) Perturbation fluid velocity profile, plotted as a function of distance from the Petri dish centre $r$, showing 
	both the radial ($u^1_r/\sin{\theta}$, black) and tangential flow ($u^1_{\theta}/\cos{\theta}$, blue) when $c = 10/45$ and $h = 8/45$.} 
	\label{fig:4}
\end{figure}

Motivated by perturbative studies of screened electrostatics near wavy boundaries \citep{Goldstein90}, we consider a small perturbation 
of the mill centre away from the middle of the Petri dish, i.e. $b = c\epsilon$ where $\epsilon \ll 1$. Expanding in powers of $\epsilon$,  
i.e. for a given function $f$ considering $f = f^0 + \epsilon f^1 + \epsilon^2 f^2 + \ldots $, (\ref{eq:streamfunctioneq}) becomes 
\begin{equation}
 \nabla^4 \varphi^{i} = \kappa^2 \nabla^2 \varphi^{i} \mbox{ for } i = 1,2,3 \ldots
\end{equation}
with corresponding boundary conditions at the edge of the Petri dish
\begin{equation}
\frac{\partial \varphi^i}{\partial r} = \frac{\partial \varphi^i}{\partial \theta} = 0 \mbox{ for } i = 1,2,3 \ldots.
\end{equation}
Furthermore, since $\Gamma$ can be expressed in polar coordinates as 
\begin{eqnarray}
\Gamma = \{ (r,\theta) \, \colon \, r = R(\theta) &=& -b \cos{\theta} + \sqrt{c^2 - b^2 \sin^2{\theta}} \nonumber\\
&=& c - \epsilon c \cos{\theta} - \epsilon^2 \frac{c \sin^2{\theta}}{2} + \mathcal{O}(\epsilon^3) \},
\end{eqnarray}
while (\ref{eq:en}) and (\ref{eq:et}) expand to become
\begin{equation}
    u_n = \frac{1}{r}\frac{\partial \varphi}{\partial \theta} 
    + \epsilon \frac{c \sin{\theta}}{r}\frac{\partial \varphi}{\partial r} 
    - \frac{\epsilon^2 c^2}{2r^2}\left( \sin{2\theta}\frac{\partial \varphi}{\partial r} 
    + \frac{\sin^2{\theta}}{r}\frac{\partial \varphi}{\partial \theta} \right) + \mathcal{O}(\epsilon^3),
\end{equation}
\begin{equation}
   u_t = -\frac{\partial \varphi}{\partial r} + \epsilon \frac{c \sin{\theta}}{r^2}\frac{\partial \varphi}{\partial \theta} 
   + \frac{\epsilon^2 c^2}{2 r^2}\left( \sin^2{\theta}\frac{\partial \varphi}{\partial r} 
   - \frac{\sin{2 \theta}}{r}\frac{\partial \varphi}{\partial \theta} \right) + \mathcal{O}(\epsilon^3),
\end{equation}
(\ref{eq:gammaun}) and (\ref{eq:gammaut}) reduce to 
\begin{eqnarray}
    \frac{1}{c}\frac{\partial \varphi^0}{\partial \theta} &+& \epsilon \left( \frac{1}{c}\frac{\partial \varphi^1}{\partial \theta} +
    \sin{\theta}\frac{\partial \varphi^0}{\partial r} + \frac{\cos{\theta}}{c}\frac{\partial \varphi^0}{\partial \theta} -
    \cos{\theta}\frac{\partial^2 \varphi^0}{\partial r \partial \theta} \right) \nonumber \\
    &&+ \epsilon^2 \Bigg{(} \frac{1}{c}\frac{\partial \varphi^2}{\partial \theta} + \sin{\theta} \frac{\partial \varphi^1}{\partial r} +
    \frac{\cos{\theta}}{c}\frac{\partial \varphi^1}{\partial \theta} - \cos{\theta}\frac{\partial^2 \varphi^1}{\partial r \partial \theta} +
    \frac{ \cos^2{\theta}}{c}\frac{\partial \varphi^0}{\partial \theta} \nonumber \\
    &&- \frac{1 + \cos^2{\theta}}{2}\frac{\partial^2 \varphi^0}{\partial r \partial \theta} - \frac{c \sin{2 \theta}}{2}\frac{\partial^2
    \varphi^0}{\partial r^2} + \frac{c \cos^2{\theta}}{2}\frac{\partial^3 \varphi^0}{\partial r^2 \partial \theta} \Bigg{)} \nonumber \\
    &+& \mathcal{O}(\epsilon^3) \Bigg{|}_{r = c} = 0. \\
    -\frac{\partial \varphi^0}{\partial r} &+& \epsilon \left( -\frac{\partial \varphi^1}{\partial r} 
    + \frac{\sin{\theta}}{c}\frac{\partial \varphi^0}{\partial \theta} 
    + c \cos{\theta}\frac{\partial^2 \varphi^0}{\partial r^2} \right) \nonumber \\
    &+& \epsilon^2 \Bigg{(} -\frac{\partial \varphi^2}{\partial r} 
    + \frac{\sin{\theta}}{c}\frac{\partial \varphi^1}{\partial \theta} 
    + c \cos{\theta}\frac{\partial^2 \varphi^1}{\partial r^2} + \frac{\sin^2 \theta}{2}\frac{\partial \varphi^0}{\partial r} 
    - \frac{\sin{2 \theta}}{2}\frac{\partial^2 \varphi^0}{\partial r \partial \theta} \nonumber \\ 
    &+& \frac{\sin{2\theta}}{2c}\frac{\partial \varphi^0}{\partial \theta} 
    + \frac{c \sin^2{\theta}}{2}\frac{\partial^2 \varphi^0}{\partial r^2} 
    - \frac{c^2 \cos^2{\theta}}{2}\frac{\partial^3 \varphi^0}{\partial r^3} \Bigg{)} \nonumber \\
    &+& \mathcal{O}(\epsilon^3) \Bigg{|}_{r = c} = c \Omega.
\end{eqnarray}
Hence, at $\mathcal{O}(1)$, namely when the circular mill is concentric with the Petri dish, we find
\begin{equation}
    \frac{\partial \varphi^0}{\partial r}\Bigg{|}_{r = 1} = \frac{\partial \varphi^0}{\partial \theta}\Bigg{|}_{r = 1} 
    = \frac{\partial \varphi^0}{\partial \theta}\Bigg{|}_{r = c} = 0 \, , \, \frac{\partial \varphi^0}{\partial r}\Bigg{|}_{r = c} 
    = -\Omega c \Longrightarrow \nonumber 
\end{equation}
\begin{equation}
    \varphi^0 = -\left(\alpha_0 K_0(\kappa r) + \beta_0 I_0(\kappa r)\right) \label{eq:varphi0eq} \Longrightarrow
\end{equation}
\begin{equation}
    u_{\theta} = c\Omega \frac{I_1(\kappa r)K_1(\kappa) - K_1(\kappa r)I_1(\kappa)}{I_1(\kappa c)K_1(\kappa) - K_1(\kappa c)I_1(\kappa)},
\end{equation}
where
\begin{subequations}
\begin{equation}
    \alpha_0 = \frac{c}{\kappa}\Omega \frac{I_1(\kappa)}{I_1(\kappa c)K_1(\kappa) - K_1(\kappa c)I_1(\kappa)},
\end{equation}
\begin{equation}
    \beta_0 = \frac{c}{\kappa}\Omega \frac{K_1(\kappa)}{I_1(\kappa c)K_1(\kappa) - K_1(\kappa c)I_1(\kappa)}.
\end{equation}
For comparison, the corresponding Couette solution is
\begin{equation}
    u_{\theta} = \frac{c^2 \Omega}{1 - c^2}\left( \frac{1}{r} - r \right).
\end{equation}
\end{subequations}
Figure \ref{fig:4}(a), plots $u_{\theta}(r)$ for both the Brinkman and Couette solutions when $c = 10/45$ and $h = 8/45$ i.e. 
the Brinkman fluid velocity field decays much faster away from the mill than the Couette fluid velocity field. 

Similarly at $\mathcal{O}(\epsilon)$, we obtain
\begin{subequations}
\begin{equation}
    \frac{\partial \varphi^1}{\partial r}\Bigg{|}_{r = 1} = 0 
    \quad , \quad \frac{\partial \varphi^1}{\partial r}\Bigg{|}_{r = c} 
    = c \cos{\theta} \frac{\partial^2 \varphi^0}{\partial r^2}\Bigg{|}_{r = c},
\end{equation}
\begin{equation}
    \frac{\partial \varphi^1}{\partial \theta}\Bigg{|}_{r = 1} = 0 \quad , \quad \frac{\partial \varphi^1}{\partial \theta}\Bigg{|}_{r = c} = - c \sin{\theta} \frac{\partial 
    \varphi^0}{\partial r}\Bigg{|}_{r = c} \Longrightarrow
\end{equation}
\end{subequations}
\begin{equation}
    \varphi^1 = - \cos{\theta}\left( \alpha_1 K_1(\kappa r) + \beta_1 I_1(\kappa r) + \gamma_1 r 
    + \frac{\delta_1}{r}  \right), \label{eq:varphi1eq}
\end{equation}
where $\{ \alpha_1 \, , \, \beta_1 \, ,\ ,  \gamma_1 \, , \, \delta_1 \}$ are known functions of $c$ and $X$ which 
satisfy the following set of simultaneous equations
\begin{subequations}
\begin{equation}
    \alpha_1 K_1(\kappa) + \beta_1 I_1(\kappa) + \gamma_1 + \delta_1 = 0,
\end{equation}
\begin{equation}
    \alpha_1 K_1(\kappa c) + \beta_1 I_1(\kappa c) + \gamma_1 c + \frac{\delta_1}{c} = \Omega c^2,
\end{equation}
\begin{equation}
    -\alpha_1\left( \kappa K_0(\kappa) + K_1(\kappa) \right) + \beta_1\left(\kappa I_0(\kappa) - I_1(\kappa) \right) + \gamma_1 - \delta_1 = 0,
\end{equation}
\begin{eqnarray}
        &-&\alpha_1\left( \frac{K_0(c/X)}{X} + \frac{K_1(c/X)}{c} \right) + \beta_1\left( \frac{I_0(c/X)}{X} 
        - \frac{I_1(c/X)}{c} \right) + \gamma_1 - \frac{\delta_1}{c^2} \nonumber \\
        &=& \kappa\Omega c^2\Bigg{(} -\frac{1}{\kappa c} 
        + \frac{I_0(\kappa c)K_1(\kappa) + K_0(\kappa c)I_1(\kappa)}{I_1(\kappa c)K_1(\kappa) - K_1(\kappa c)I_1(\kappa)} \Bigg{)}.
\end{eqnarray}
\end{subequations}
Figure \ref{fig:4}(b) plots $u^1_r/\sin{\theta}$ and $u^1_{\theta}/\cos{\theta}$ as functions of $r$ when $c = 10/45$ and $h = 8/45$. This perturbation flows also decays exponentially away from the mill i.e. the Brinkman term still plays a key role. As will be shown in \S\ref{nearfieldcircularmill}, this perturbation flow leads to the centre of the mill drifting clockwise on a circle centered the middle of the Petri dish, namely the stationary point when the mill and the Petri dish are concentric is unstable.
 \subsection{Far-field Solution} \label{farfieldpeturb}

\begin{figure}
	\centering
	\includegraphics[trim={0 0cm 0 0cm}, clip, width=\textwidth]{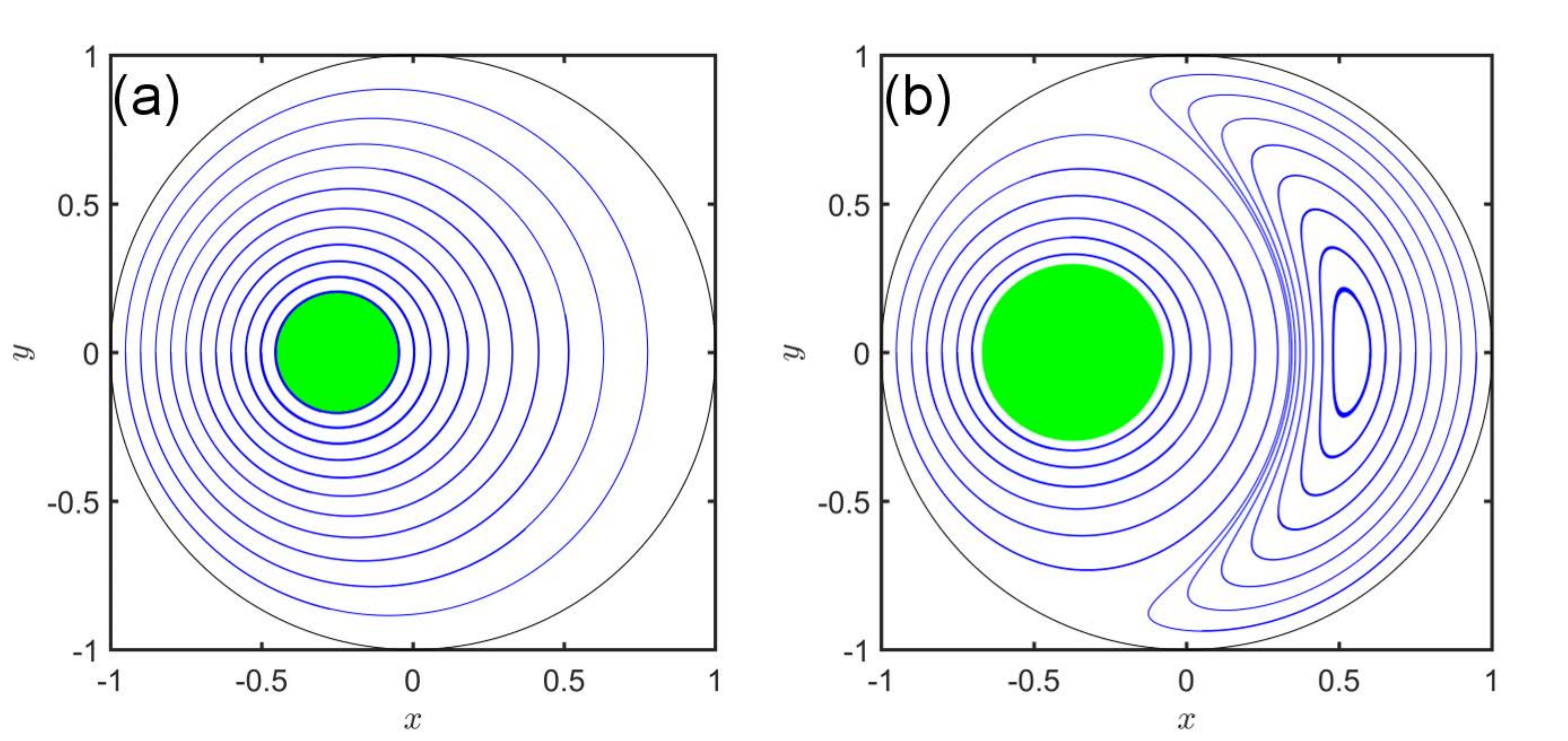}
	\caption{Streamlines of the flow highlighting the two distinct possibilities; namely no stagnation points in (a) 
	where $b = 0.25$ and $c = 0.2$ and stagnation points in (b) where $b = 0.373$ and $c = 0.298$.}
	\label{fig:5}
\end{figure}

When the circular mill is away from the centre of the Petri dish ($b = \mathcal{O}(1)$), the boundary conditions at the edge of 
the mill can no longer be expressed straightforwardly in terms of the polar coordinates $(r,\theta)$. Instead we switch to the 
bipolar coordinates $(\eta , \xi)$ utilising the transformations
\begin{equation}
x = a \frac{\sinh{\eta}}{\cosh{\eta} - \\cos{\xi}} + d \mbox{ and } y = a \frac{\\sin{\xi}}{\cosh{\eta} - \\cos{\xi}}.
\end{equation} 
In particular, we can map the outer boundary, $r = 1$, to $\eta = \eta_1$ and the disc boundary to $\eta = \eta_2$ by defining 
the constants $a,d,\eta_1$ and $\eta_2$ as satisfying
\begin{equation}
d = -a \coth{\eta_1} \, , \, b = a(\coth{\eta_1}  - \coth{\eta_2}) \, , \, 1 = \frac{a}{\sinh{\eta_1}} \mbox{ and } c = \frac{a}{\sinh{\eta_2}}.
\end{equation}
i.e.
\begin{equation}
\eta_1 = \ln{\left( a + \sqrt{a^2 + 1} \right)} \, , \, \eta_2 = \ln{\left( \frac{a + \sqrt{a^2 + c^2}}{c} \right)},
\end{equation}
\begin{equation} 
a = \frac{1}{2b}\left(\sqrt{(1 + c^2 - b^2)^2 - 4c^2}\right).
\end{equation}
In this basis, the system becomes
\begin{equation}
    \nabla^2 \left( \nabla^2 - \kappa^2 \right)\varphi = 0 \mbox{ where } \nabla^2 
    = \frac{1}{h^2}\left( \frac{\partial^2}{\partial \xi^2} + \frac{\partial}{\partial \eta^2} \right) \mbox{ and } 
    h = \frac{a}{\cosh{\eta} - \cos{\xi}}, \label{eq:startingbipolarstreamfunction}
\end{equation}
with boundary conditions
\begin{equation}
\frac{\partial \varphi}{\partial \xi} = \left\{
\begin{array}{ll}
0, & \eta = \eta_1 \\[2pt]
0,         & \eta = \eta_2.
\end{array} \right. \mbox{ and } 
\frac{1}{h}\frac{\partial \varphi}{\partial \eta} = \left\{
\begin{array}{ll}
0, & \eta = \eta_1 \\[2pt]
-\Omega c,         & \eta = \eta_2.
\end{array} \right.
\end{equation}
Now in general this does not admit an analytic solution. However for large mills away from the Petri dish centre, 
namely $1/a > \kappa $, the biharmonic term dominates and thus using \cite{Melesko99}, (\ref{eq:startingbipolarstreamfunction}) 
reduces to
\begin{equation}
    \nabla^4 \varphi = 0 \longrightarrow \frac{\partial^4 \Phi}{\partial \xi ^4} + 2 \frac{\partial^4 \Phi}{\partial \xi^2 \partial \eta^2} + \frac{\partial^4 \Phi}{\partial \eta^4} + 2 \frac{\partial^2 \Phi}{\partial \xi^2} - 2 \frac{\partial^2 \Phi}{\partial \eta^2} + \Phi = 0 \mbox{ where } \Phi = \frac{\varphi}{h}.
\end{equation}
From \cite{Kazakova15}, this yields the analytic solution
\begin{equation}
\varphi = N(\eta) + \frac{M(\eta)}{\cosh(\eta) - \\cos(\xi)}, \label{eq:bipolarstreamfunction}
\end{equation}
where
\[
N(\eta) = A\eta - F\cosh{2\eta} - G\sinh{2\eta}, 
\]
\[
M(\eta) = B\sinh{\eta} + C\cosh{\eta} + E\eta\sinh{\eta} + F\cosh{\eta}\cosh{2\eta} + G\cosh{\eta}\sinh{2\eta}.
\]
Here A, B, C, E, F and G are constants which, letting $\alpha = \eta_1 + \eta_2$ and $\beta = \eta_1 - \eta_2$, satisfy
\[
A = \frac{\Omega c a \cosh{\beta}}{\sinh{\beta}}\frac{\beta \sinh{\eta_2} 
- \sinh{\beta} \sinh{\eta_1}}{\beta (\cosh{\alpha}\cosh{\beta} - 1) - \sinh{\beta} (\cosh{\alpha} - \cosh{\beta})},
\]
\[
E = \Omega c a \frac{\cosh{\beta} \cosh{\eta_1} - \cosh{\eta_2}}{\beta (\cosh{\alpha}\cosh{\beta} - 1) 
- \sinh{\beta} (\cosh{\alpha} - \cosh{\beta})},
\]
\[
C = \frac{\sinh{\eta_2}}{2}(E \sinh{\eta_2} + A \cosh{\eta_2} + \Omega c a) + \frac{\sinh{\eta_1}}{2}(E \sinh{\eta_1} + A \cosh{\eta_1}),
\]
\[
B = -E \eta_2 - \cosh{\eta_2}(E \sinh{\eta_2} + A \cosh{\eta_2} + \Omega c a),
\]
\[
F = -\frac{A}{2}\frac{\sinh{\alpha}}{\cosh{\beta}} \, , \, G = \frac{A}{2}\frac{\cosh{\alpha}}{\cosh{\beta}}.
\]
From \cite{Kazakova15}, this flow field takes one of two forms. When the mill is relatively close to the centre of the Petri dish, 
the flow has no stagnation points and the streamlines are circular (Figure \ref{fig:5}a gives a typical example). When the mill is 
close to the boundary of the Petri dish, the flow has a stagnation point (Figure \ref{fig:5}b gives a typical example). 
Mathematically, a stagnation point exists when $u_{\eta} = u_{\xi} = 0$ i.e. $\xi = 0$ and $\eta = \eta^{\star}$ where 
$\eta_1 < \eta^{\star} < \eta_2$ satisfies
\begin{equation}
V(\eta^{\star}) = 0 \mbox{ : } V(\eta) = \frac{d}{d \eta}\left( N(\eta) + \frac{M(\eta)}{\cosh{\eta} - 1} \right).
\end{equation} 
Note that when $\xi = \pi$, although $u_{\eta} = 0$, $u_{\xi} \neq 0 \, \forall \, \eta \in (\eta_1, \eta_2)$. Without 
loss of generality, let $\Omega > 0$ i.e. $V(\eta_1) = 0$ while $V(\eta_2) <0$. If $V'(\eta_1) < 0$, V decreases monotonically 
and no such $\eta \in (\eta_1, \eta_2)$ exists. Conversely if $V'(\eta_1) > 0$, V achieves positive values in 
$[\eta_1, \, \eta_2 ]$ and so by the intermediate value theorem, such a $\eta \in (\eta_1, \eta_2)$ exists. Hence 
in $\{ b, \, c\}$ phase space, the critical curve separating the two regions satisfies $V'(\eta_1) = 0$. Furthermore 
from \cite{Kazakova15}, a good approximation to the boundary is the interpolation curve
\begin{equation}
b(c) = 0.36(1 - c) + 0.08 c(c-1).
\end{equation} 

\section{The Drift of the Circular Mill Centre} \label{millcentredrift}

The flow exerts a force $\boldsymbol{F}$ on the mill where $\boldsymbol{F} = F_x  \boldsymbol{\hat{x}} + F_y \boldsymbol{\hat{y}} $ satisfies
\begin{equation}
\boldsymbol{F} = \int^{\pi}_{-\pi} (\sigma_{\eta\eta} \boldsymbol{\hat{\eta}} + \sigma_{\eta\xi} 
\boldsymbol{\hat{\xi}})_{\eta = \eta_2} \, h \, d\xi. \label{eq:viscousforce}
\end{equation}
Here the bipolar basis vectors $\boldsymbol{\hat{\eta}}$ and $\boldsymbol{\hat{\xi}}$ satisfy $\boldsymbol{\hat{\eta}} = 
f \boldsymbol{\hat{x}} + g \boldsymbol{\hat{y}}$ and $\boldsymbol{\hat{\xi}} = g \boldsymbol{\hat{x}} - f \boldsymbol{\hat{y}}$ where
\begin{equation}
f = \frac{1 - \cosh{\eta}\,\cos{\xi}}{\cosh{\eta} - \cos{\xi}} \mbox{ and } g = - \frac{\sin{\xi}\sinh{\eta}}{\cosh{\eta} - \cos{\xi}},
\end{equation}
while $\sigma_{\eta\eta}$ and $\sigma_{\eta\xi}$ are components of the stress tensor $\sigma_{ij} = -p \delta_{ij} 
+ \mu \left(\partial u_i/\partial x_j + \partial u_j/\partial x_i\right)$. Since this is not a standard result given 
in the literature (\cite{Wakiya75} is the closest reference which can be found), for completeness appendix \ref{hopefullycorrect} 
gives the full form of $\partial u_i/\partial x_j$ when expressed in bipolar coordinates for general $\boldsymbol{u}$.

This system, in a domain symmetric about the line $\theta = 0$, is forced by a fluid flow 
even in $\theta$. Hence, since it admits a general separable form where each term is either even or odd in 
$\theta$ (\ref{eq:separableform}), $\varphi$ is even in $\theta$ and hence $p$ and $\sigma_{rr}$ are also even in $\theta$. Similarly, $\sigma_{r \theta}$ is odd in $\theta$ and hence from rewriting (\ref{eq:viscousforce}) in terms of cylindrical polar coordinates, $F_x = 0$. This force causes the mill 
centre to slowly drift on a larger timescale than the period of rotation of a mill, maintaining a constant distance 
from the centre of the Petri dish.

In general, $F_y$ does not admit an analytic form. However, as in $\S$\ref{nearfieldpeturb} and $\S $\ref{farfieldpeturb}, 
further progress can be made analytically for circular mills both close to and far away from the centre of the Petri dish.  

\subsection{Near-field Circular Mill} \label{nearfieldcircularmill}
Building from $\S$\ref{nearfieldpeturb}, substituting (\ref{eq:varphi0eq}) and (\ref{eq:varphi1eq}) 
into (\ref{eq:verticalaveragedstokesflow}) using standard properties of modified Bessel functions and then integrating yields
\begin{equation}
    p^0 = p_0 \quad , \quad p^1 = \mu \kappa^2\sin{\theta}\left( \frac{\delta_1}{r} - \gamma_1 r \right),
\end{equation}
where $p_0$ is a constant. Furthermore, since $\sigma_{rr} = -p + 2\mu \, {\partial u_r}/{\partial r}$ while $\sigma_{r\theta} 
= \mu({\partial u_{\theta}}/{\partial r} + ({\partial u_r}/{\partial \theta})/r - u_{\theta}/r)$, we obtain
\begin{equation}
    \sigma^0_{rr} = 0 \quad , \quad \sigma^0_{r\theta} = \alpha_0 \mu \left( \kappa^2K_0(\kappa r) 
    + \frac{2\kappa K_1(\kappa r)}{r} \right) + \beta_0 \mu \left( \kappa^2 I_0(\kappa r) - \frac{2\kappa I_1(r/X)}{r} \right),
\end{equation}
\begin{eqnarray}
    \sigma^1_{rr} = \mu \sin{\theta} && \left( -2\alpha_1 \left( \frac{\kappa K_0(\kappa r)}{r} + \frac{2K_1(r/X)}{r^2} \right) 
    + 2\beta_1 \left( \frac{\kappa I_0(\kappa r)}{r} -\frac{2 I_1(\kappa r)}{r^2} \right) \nonumber \right. \\
    &&+ \left. \gamma_1 \kappa^2 r - \delta_1 \left( \frac{\kappa^2}{r} + \frac{4}{r^3} \right) \right),
\end{eqnarray}
\begin{eqnarray}
    \sigma^1_{r\theta} = \mu \cos{\theta} && \left( \alpha_1 \left( \frac{2\kappa K_0(\kappa r)}{r} + K_1(\kappa r) 
    \left( \kappa^2 + \frac{4}{r^2} \right) \right) \nonumber \right. \\ 
    &&- \left. \beta_1 \left( \frac{2\kappa I_0(\kappa r)}{r} - I_1(\kappa r) \left( \kappa^2 
    + \frac{4}{r^2} \right) \right)+ \frac{4\delta_1}{r^3} \right).
\end{eqnarray}
Hence, the flow exerts a force $\boldsymbol{F}$ on the mill, where $\boldsymbol{F} = F_x \boldsymbol{\hat{x}} 
+ F_y \boldsymbol{\hat{y}}$ satisfies
\begin{equation}
    \boldsymbol{F} = \int^{\pi}_{\pi} \left( \sigma_{rr} \boldsymbol{\hat{r}} 
    + \sigma_{r\theta} \boldsymbol{\hat{\theta}} \right) \Bigg{|}_{r = c} c d\theta \Longrightarrow
\end{equation}
\begin{equation}
F^0_x = F^0_y = F^1_x = 0, 
\end{equation}
\begin{equation}
F^1_y = \pi\mu\kappa^2 c \left( \alpha_1 K_1(\kappa c) + \beta_1 I_1(\kappa c) + \gamma_1 c - \frac{\delta_1}{c}\right) 
= \pi \mu\kappa^2 c^2\left( \Omega c - \frac{2\delta_1}{c^2} \right).
\end{equation}
Note that out the front of this expression, we have $(\kappa c)^2$ rather than $c^2$ i.e. the effective radius of the mill is modulated by the screening length $\kappa$. For the values taken in figure \ref{fig:4}, $F^1_y$ is positive, i.e. the mill centre drifts clockwise in a circle centered the middle of the Petri dish. 

\subsection{Far-field Circular Mill} \label{farfieldforce}
Since (\ref{eq:verticalaveragedstokesflow}) reduces in this case to the Stokes equations, $F_x = 0$ follows 
immediately by utilising the properties of a Stokes flow. Reversing time and then reflecting in the $x$ axis returns 
back to the original geometry but with the sign of $F_x$ flipped i.e. $F_x = - F_x \rightarrow F_x = 0$. 
Substituting (\ref{eq:bipolarstreamfunction}) into (\ref{eq:verticalaveragedstokesflow}) and then integrating gives the pressure
\begin{eqnarray}
p = \frac{2\mu}{a^2}(&E&\sinh{\eta} \, \sin{\xi} + F\sinh{2\eta} \, \sin{2\xi} - 2F\sinh{\eta} \, \sin{\xi} \nonumber \\
&+& G\cosh{2\eta} \, \sin{2\xi} - 2G\cosh{\eta} \, \sin{\xi}).
\end{eqnarray} 
Shifting the basis vectors back to Cartesian coordinates, the force can be expressed in the form
\begin{equation}
\boldsymbol{F} = \frac{\mu}{a^2}\int^{\pi}_{-\pi} (f_x \boldsymbol{\hat{x}} + f_y \boldsymbol{\hat{y}})_{\eta = \eta_2}  \, h \, d\xi,
\end{equation}
where $f_x$ and $f_y$ are explicit functions of $\eta$, $\xi$ and $\{ A,B,C,E,F,G\}$. However, $f_x$ is odd with respect to $\xi$ at $\eta = \eta_2$ since
\begin{eqnarray}
&&\frac{1}{2}\left(f_x(\eta,\xi) + f_x(\eta,-\xi)\right) = \nonumber \\
\frac{2\sin^2{\xi}(1-\cosh{\eta}\, \cos{\xi})}{(\cosh{\eta} - \cos{\xi})^2}(&&C\cosh{\eta} + B\sinh{\eta} + E\eta\sinh{\eta} + F\cosh{\eta}(2\cosh^2{\eta} - 1) \nonumber \\
&&+ 2G\cosh^2{\eta}\sinh{\eta}) = 0 \mbox{ at } \eta = \eta_2.
\end{eqnarray}
Therefore as expected $F_x = 0$. $f_y$ can be similarly simplified, removing the terms odd in $\xi$, to give
\begin{equation}
\boldsymbol{F} = \frac{\mu \boldsymbol{\hat{y}}}{8 a} \sum_{i=0}^{4} g_i(\eta_2) I_{i}, \label{finalforcevalue}
\end{equation}
where $g_i = g_i(\eta) \, : i \in \{0,1,2,3,4\}$ are given for completeness in Appendix \ref{algbera} while $I_n$ satisfies
\begin{equation}
I_n = \int^{\pi}_{\pi} \frac{\cos{(n\,\xi)}}{(\cosh{\eta} - \cos{\xi})^3} d\xi.
\end{equation}
\begin{figure}
	\centering
	\includegraphics[trim={0 0cm 0 0cm}, clip, width=\textwidth]{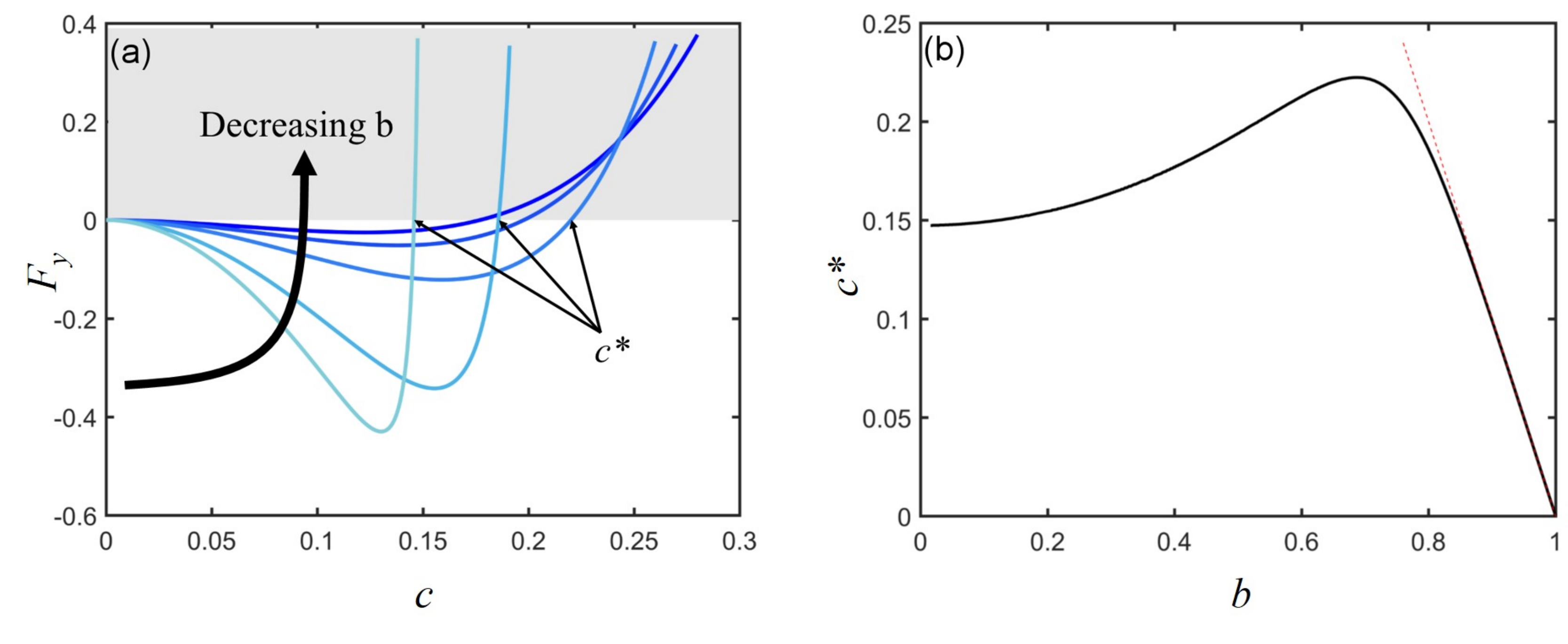}
	\caption{(a) The force that the flow exerts on the disc expressed as a function of c for a range of values of $b$ $(0.85, 0.8, 0.65, 0.52, 0.4)$ (b) The critical radius $c^{\star}$ expressed as a function of $b$. As $b \rightarrow 1$, $c^{\star} \rightarrow 1 - b$ (the red dashed line). }
	\label{fig:6}
\end{figure}  
To investigate this force more quantitatively, we numerically calculate $F_{y}$ from (\ref{finalforcevalue}) as a function of $b$ and $c$ by utilising MATLAB's symbolic variable toolbox. Figure \ref{fig:6}a plots $F_{y}$ as a function of $c$ for a range of values for $b$ (chosen to demonstrate the full phase space of behaviour of $F_{y}(b,c)$). Large mills have positive $F_y$ (in the grey region), i.e. the mill centre drifts in the same angular direction as the worms. Small mills have negative $F_y$ (in the white region), i.e. the mill centre drifts in the opposite direction to the worms. 

Hence, we define the critical radius $c^{\star} = c^{\star}(b)$ as the mill radius at which $F_y = 0$ (plotted as a function of $b$ in Figure \ref{fig:6}b). Note that when $b$ is large the critical geometry is a lubrication flow since $b + c^{\star} \rightarrow 1$. Furthermore, $c^{\star}$ has a maximum of 0.222 at $b = 0.70$.

\subsection{Comparison with Experiments}

\begin{figure}
	\centering
	\includegraphics[trim={0 0cm 0 0cm}, clip, width=\textwidth]{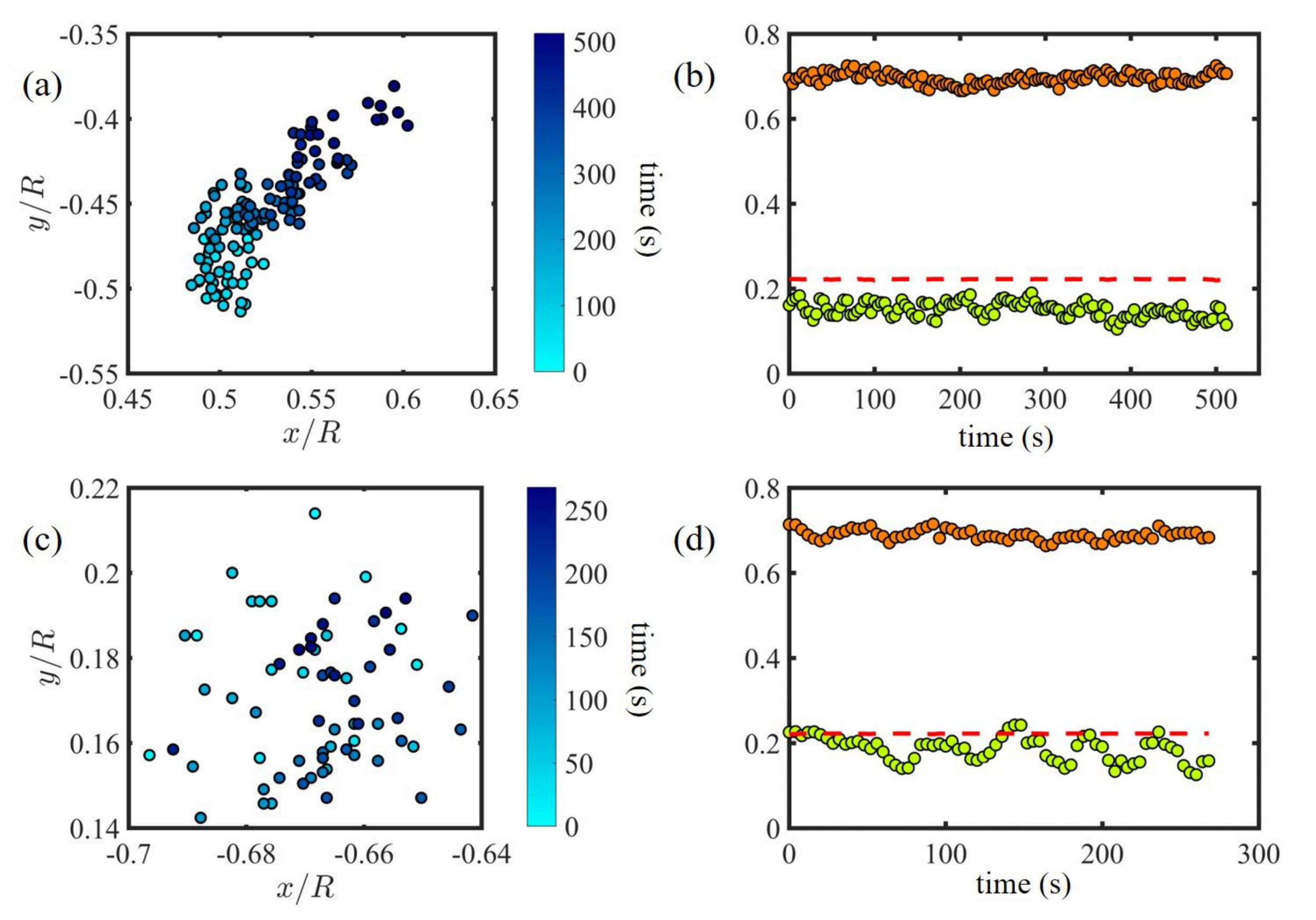}
	\caption{Circular mill data for two different experiments [(a) + (b) and (c) + (d)]. 
		(a) and (c) : The location of the mill centre plotted onto the ($x$,$y$) plane.
		Points shaded a darker blue denote a later time, as quantified by the colorbar.
		(b) and (d) : $b$ (orange filled circles), $c$ (green filled circles) and the critical radius $c^{\star}(b)$ (red dashed line),   plotted as functions of time.
		Here, $\{ x, \, y, \, b,  \, c, \, c^{\star} \}$ have all been normalised by the Petri dish radius (denoted $R$).}
	\label{fig:7}
\end{figure}

We now compare these predictions with experimental data for $b(t)$ and $c(t)$, generated using the methodology given in \S\ref{millingexperiments}.
Unlike the simple circle considered in the model, the shape of a real circular mill is complicated. Not only does a mill at any one time consist of thousands of individual worms but also, as the mill evolves, this population changes as worms enter and leave. Hence, mills typically have constantly varying effective radii and are not simply connected. Furthermore, the edges of a circular mill are not well-defined, leading to a greater uncertainty in measuring the mill radius.  
However, despite these complications, the experimental results agree well with the predictions made above in \S\ref{millcentredrift}. Within experimental uncertainty, $b$ is constant i.e. the centre of the mills do indeed drift on circles centred at the middle of the arena. Furthermore, the direction of drift also matches with the theory given in \S\ref{farfieldforce} for the force on a mill in the far field. 

To illustrate this, consider Figure \ref{fig:7} which presents graphically the experimental data for two representative experiments which sit at either end of the phase space of mill centre trajectories. In the first experiment (Figures \ref{fig:7}(a) and \ref{fig:7}(b)), the circular mill radius $c$ decays slowly over time, always being less than the critical radius $c^{\star}(b)$ (in Figure \ref{fig:7}(b) the green points lie below the red dashed line). Hence we are in the white region of the phase space in Figure \ref{fig:6}. Since the worms are moving clockwise, the model predicts that the mill centre should drift anti-clockwise, increasing in angular speed as time progresses. This is indeed what we see in Figure \ref{fig:7}(a) with the darker later time blue points less clustered together than the lighter earlier time points.   

In contrast, we see much more variation in $c$ in the second experiment (Figures \ref{fig:7}(c) and \ref{fig:7}(d)) with points both above and below $c^{\star}$ (in Figure \ref{fig:7}(d) green points lie either side of the red dashed line). The predicted sign of $F_y$ thus oscillates i.e. the model predicts that the net drift of the mill centre should be minimal. This is indeed what we see in Figure \ref{fig:7}(c) with light and dark blue points equally scattered.   

\section{Binary Circular Mill Systems} \label{binarystagnation}

\begin{figure}[t]
	\centering
	\includegraphics[trim={0 0cm 0 0cm}, clip, width=\textwidth]{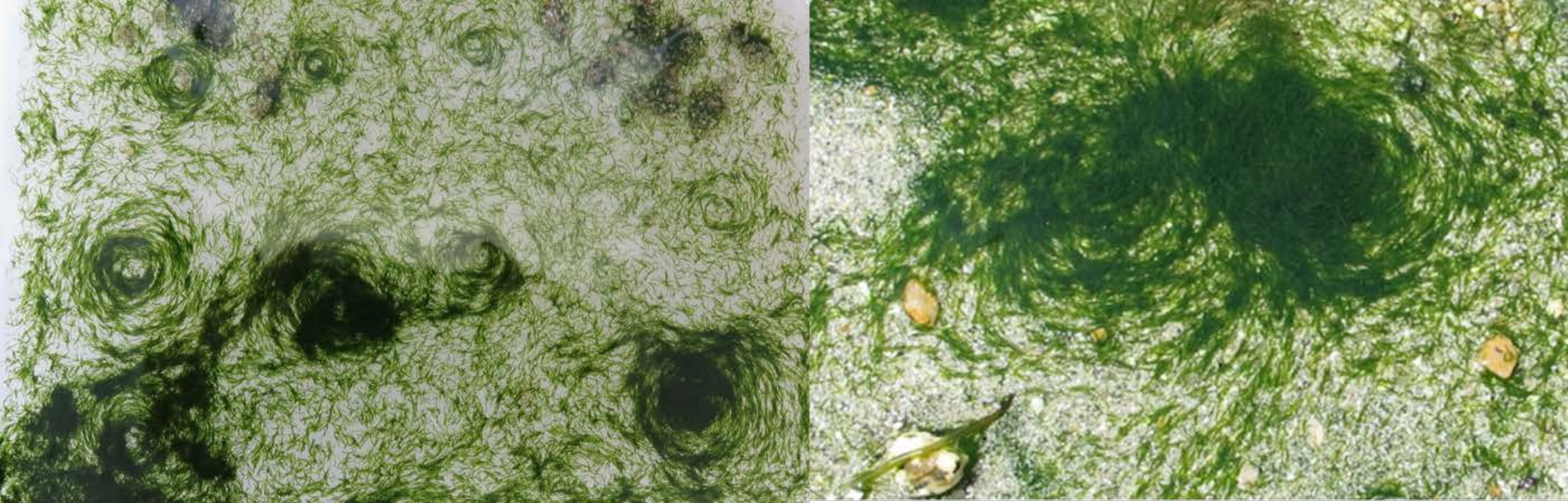}
	\caption{Systems containing multiple mills: (a) Nine circular mills of different sizes observed \textit{ex situ} (in a tub), (b) Binary circular mill system observed \textit{in situ} (on a beach). Both images are reproduced from \cite{SendovaFranks18}.}
	\label{fig:8}
\end{figure}

During the evolution of the system, multiple mills can emerge at the same time (Figure \ref{fig:8} and Supplementary Video 2). This is to be expected since the worms can only interact locally with each other and hence can not coordinate globally to produce a single mill. Here for simplicity we will only consider the most common example of this phenomenon, namely a pair of circular mills. Since the radii of the mills is of the same order of magnitude as the distance between them, a perturbation expansion in terms of c is not possible. Hence the method utilised in $\S$\ref{millingmaths} can not yield an analytic solution here. 

However, using the insight revealed from $\S$\ref{millingmaths} regarding the flow field produced by an individual mill, we can explain the experimentally observed behaviour from a fluid dynamical viewpoint. In particular, we can explain both the location where the second mill forms and the direction in which it rotates and predict the stability of the binary system. Experimentally, we observed a total of nine binary circular mill systems (summarised in appendix \ref{millingexperimentsdata}). Figure \ref{fig:9}(a-c) gives a snapshot from three of these experiments. Using the theoretical model, streamlines for the flow produced by each of the two mills if they existed in isolation were generated and superimposed on the same plot (Figures \ref{fig:9}(d-f)). 

The first important observation is that secondary mills only appear when the flow produced by the first mill has a stagnation point, forming in the corresponding stagnation point region. All nine observed binary circular mill systems obey this hypothesis while all observed circular mills which do not generate a stagnation point are stable to the emergence of secondary mills. Note that this relation is not a one-to-one correspondence between having stagnation points and secondary mills emerging. Many other factors can prevent secondary mills forming e.g. a low density of worms swimming in the stagnation point region.

Furthermore, the worms forming the secondary mill tend to swim in the direction of the flow around the stagnation point i.e. the two mills tend to rotate in the same angular direction. In seven of the nine binary circular mill systems examined, the mills rotate in the same direction while the second mill in one of the other systems is seeded by a single flotilla of worms who were tracking around the edge of the Petri dish. 

Finally, we can gain a qualitative understanding of the stability of the binary system from looking at the streamlines produced by the second mill. If these streamlines do not have a stagnation point in the vicinity of the first mill (Figures \ref{fig:9}(a) and \ref{fig:9}(d)), the system is unstable as the first mill breaks up.  Alternatively if a stagnation point exists and aligns with the first mill (Figures \ref{fig:9}(b) and \ref{fig:9}(e)), the system will be stable. Figures \ref{fig:9}(c) and \ref{fig:9}(f) show the intermediate regime where the first mill is partly (but not fully) inside the second mill's stagnation region. The system is unstable over a much longer time scale. In this particular case, since the first mill is much larger than the second mill, it dominates and the second mill breaks up.

\begin{figure}
	\centering
	\includegraphics[trim={0 0cm 0 0cm}, clip, width=\textwidth]{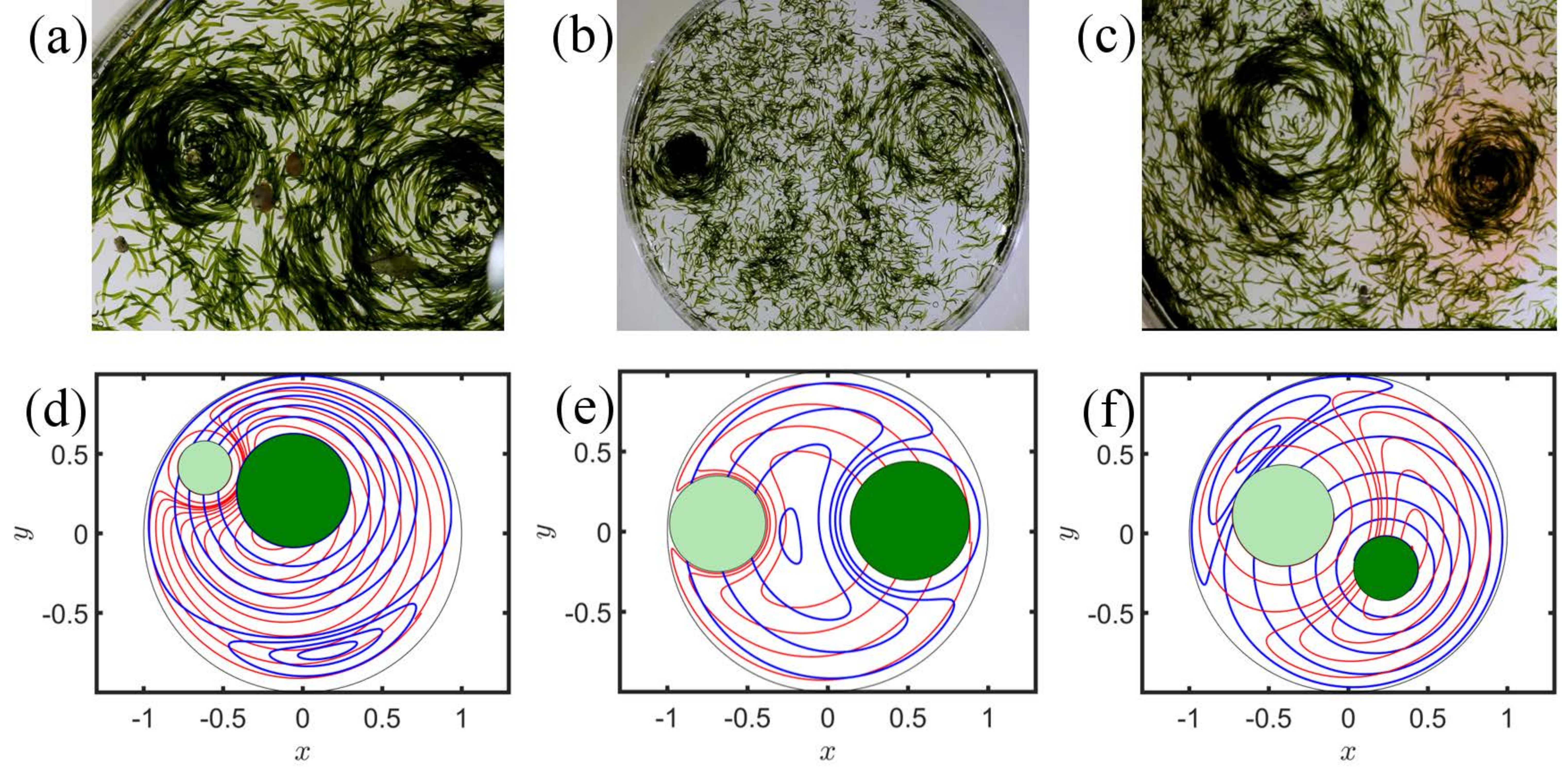}
	\caption{Snapshots of a binary circular mill system for three distinct experiments together with corresponding streamline plots (First mill is light green with red streamlines, second mill is dark green with blue streamlines): (a) Unstable with the second mill dominating, (b) Stable, (c) Unstable on a longer timescale with the first mill dominating.}
	\label{fig:9}
\end{figure}

\section{Milling Conclusions}

Vortex motions in animal groups have been studied for over a century in many animal species. In this paper, we have demonstrated for the very first time that in order to understand these behaviours in aquatic environments, of which the circular milling of $\textit{S. roscoffensis}$ is a prime example, one has to understand the underlying fluid dynamics of the system. From the drift of the vortex centre to the formation of secondary vortices and their subsequent stability, it is fundamentally the fluid flow processes that drive these mesmerising and constantly evolving structures. This fluid velocity field may allow nutrient circulation as well as providing an efficient method of dispersal of waste products away from the main body of worms. Furthermore, it exerts a force on the circular mill which causes the mill to slowly drift. In particular, for a single mill in a circular arena, the centre of the mill drifts on a circle whose centre is the middle of the arena. 

We present a simple model for the system, (a rigid disc rotating in a Stokes flow), parametrised by only two key variables; namely the radius of the mill $c$ and the distance to the centre of the arena $b$. This fits the experimental results well, both in terms of the mill centre drift direction but also the predicted streamlines. Utilising this understanding, we are able to shed light on the fluid dynamical stability of circular mills. Secondary circular mills form around stagnation points of the flow. The resulting system evolves to one of two kinds of stable states; namely a single mill with no nearby stagnation points or a set of linked mills where each mill centre is located in the stagnation region of another mills.  Although in real life the geometry of the arena is more complicated than our circular model, the same principle remains, namely that stagnation points of the flow occur near a mill when that mill is close to a boundary. This allows the worm population to passively organise towards the arena centre without needing to know the exact extent of the domain. Typically the arena centre will be less shaded and more resource rich.

A next step is to estimate the speed of drift of the mill centre.  As each worm secretes a layer of mucus around itself, creating a non-Newtonian boundary layer between the mill and the bottom of the Petri dish, 
both the thickness of this boundary layer and the mechanical properties of the mucus need to be quantified before 
one is able to calculate this drift. A second line of enquiry results from the fact that a dense core of stationary worms is often experimentally observed to form in the centre of a mill. This core can be unstable and break up or it can take over the whole mill, forming a biofilm.
At a more microscopic level, it is of interest to examine the extent to which the formation and breakup of
mills can be captured by the kinds of continuum models that have been used successfully to study collective
behaviour in bacterial systems \citep{SaintillanShelley}, where vortex formation is now
well-established \citep{Wioland}.

\backsection[Funding]{This work was supported in part by the Engineering and Physical Sciences Research 
Council, through a doctoral training fellowship (GTF) and an
Established Career Fellowship EP/M017982/1 (REG), an ERC Consolidator grant 682754 (EL), and 
by the Schlumberger Chair Fund (REG).}

\backsection[Declaration of interests]{The authors report 
no conflict of interest.}

\backsection[Author ORCID]{G.T. Fortune, https://orcid.org/0000-0003-0817-9271; A. Worley, https://orcid.org/0000-0002-7734-7841; A.B. Sendova-Franks, 
https://orcid.org/0000-0001-9300-6986; N.R. Franks, https://orcid.org/0000-0001-8139-9604; 
K.C. Leptos, \\ https://orcid.org/0000-0001-9438-0099; E. Lauga, https://orcid.org/0000-0002-8916-2545;
\\ R.E. Goldstein, https://orcid.org/0000-0003-2645-0598}

\appendix
\section{Experimental Data on Binary Circular Mill Systems} \label{millingexperimentsdata}
Table \ref{table:1} gives collocated experimental data of the evolution of circular mills across eighteen distinct experiments. The experimental net drift was obtained by plotting the angle between the line through the centres of the mill and Petri dish and a fixed reference line as a function of time. Linearly interpolating this data, if the magnitude of the gradient of the plotted line is greater than $1.05$ rad/h (i.e. changes by more than 10$^{\circ}$ during a experiment of typical duration $10$ minutes), then we can definitively say that there is a net drift i.e. clockwise if the gradient of the line is negative or anticlockwise if the gradient of the line is positive. Otherwise we write none, since there is no observable net drift within the bounds of experimental error. Similarly, the net drift predicted by net drift was obtained by plotting $c - c^{\star}(b)$ as a function of time. If the mean of these data points is greater than one standard deviation, then we predict that the mill should drift clockwise. If the mean is less than zero but has magnitude greater than on standard deviation, we predict that the mill should drift anticlockwise. Otherwise, we predict that there is should be no observable net drift within the bounds of experimental error.

While the radius of a circular mill $c$ varies considerably throughout its evolution, its centre remains within experimental error at a constant distance $b$ from the centre of the Petri dish. For fourteen of the eighteen mills, the predicted net direction of drift of the mill centre from the model (assuming that the mill centre drifts in the direction of the force that the flow imposes onto it) matches with the actual net direction. The discrepancy in the other four experiments arises from inertial effects, which particularly come into play for circular mills close to the centre of the Petri dish (experiments 8 and 9) where $\boldsymbol{F}$ from (\ref{finalforcevalue}) is small. Table \ref{table:2} gives the corresponding data for nine distinct binary circular mill systems.   
\begin{table}
	\begin{center}
		\footnotesize
		\begin{tabular}{lccccccc}
			Exp. & Variation & Variation & Mill & Experimental  & Net Drift & Stag. & Leads to \\
			\#  & in $b$ & in $c$ & Orient. & Net Drift & Predicted & Points & Binary\\
			&(Min-Max)&(Min-Max)&&&By Theory&&System\\
			&&&&&&&\\
            1 & 0.51 - 0.63 & 0.09 - 0.27 & CW & None & None & $\checkmark$ & $\checkmark$\\ 
            2 & 0.66 - 0.71 & 0.13 - 0.24 & CW & None & ACW & $\checkmark$ & $\checkmark$\\
            3 & 0.42 - 0.49 & 0.14 - 0.28 & CW & None & None & $\checkmark$ & $\boldsymbol{?}$\\
            4 & 0.67 - 0.73 & 0.10 - 0.19 & CW & ACW & ACW & $\checkmark$ & $\boldsymbol{?}$\\
            5 & 0.14 - 0.20 & 0.16 - 0.25 & CW & CW & CW & $\times$ & $\times$\\
            6 & 0.62 - 0.69 & 0.14 - 0.23 & CW & None & ACW & $\checkmark$ & $\boldsymbol{?}$\\
            7 & 0.12 - 0.20 & 0.18 - 0.31 & CW & CW & CW & $\times$ & $\times$\\
            8 & 0.07 - 0.18 & 0.15 - 0.32 & CW & ACW & CW & $\times$ & $\times$\\
            9 & 0.07 - 0.12 & 0.22 - 0.32 & CW & None & CW & $\times$ & $\times$\\
            10 & 0.22 - 0.26 & 0.12 - 0.27 & CW & None & None & $\times$ & $\times$\\
            11 & 0.21 - 0.29 & 0.17 - 0.41 & CW & CW & CW & $\times$ & $\times$\\
            12 & 0.23 - 0.30 & 0.17 - 0.32 & CW & CW & CW & $\times$ & $\times$\\
            13 & 0.31 - 0.34 & 0.23 - 0.31 & CW & CW & CW & $\checkmark$ & $\boldsymbol{?}$\\
            14 & 0.46 - 0.54 & 0.23 - 0.40 & CW & CW & CW & $\checkmark$ & $\boldsymbol{?}$\\
            15 & 0.35 - 0.41 & 0.27 - 0.35 & CW & CW & CW & $\checkmark$ & $\boldsymbol{?}$\\
            16 & 0.57 - 0.64 & 0.20 - 0.32 & CW & CW & CW & $\checkmark$ & $\times$\\
            17 & 0.26 - 0.36 & 0.19 - 0.40 & CW & CW & CW & $\checkmark$ & $\boldsymbol{?}$\\
            18 & 0.25 - 0.30 & 0.19 - 0.34 & CW & CW & CW & $\checkmark$ & $\boldsymbol{?}$\\
		\end{tabular}
		\normalsize
		\caption{Radius $c$, distance from the arena centre $b$, and orientation data from the evolution of eighteen distinct circular mills. CW is clockwise while ACW is anticlockwise.}
	     \label{table:1}
\end{center}
\end{table}
\begin{table}
	\begin{center}
		\small
		\begin{tabular}{lcccccc}
			Experiment & Variation & Variation & First Mill & Variation & Variation & Second Mill \\
			\quad \quad \#  & in $b_1$ & in $c_1$ & Orientation & in $b_2$ & in $c_2$ & Orientation \\
			&(Min-Max)&(Min-Max)&&(Min-Max)&(Min-Max)&\\
			&&&&&&\\
			\quad \quad 1 & 0.69 - 0.75 & 0.14 - 0.24 & CW & 0.38 - 0.42 & 0.15 - 0.22 & CW \\
			\quad \quad 2 & 0.72 - 0.78 & 0.10 - 0.22 & CW & 0.27 - 0.34 & 0.19 - 0.36 & CW\\
			\quad \quad 3 & 0.35 - 0.39 & 0.09 - 0.15 & CW & 0.05 - 0.11 & 0.21 - 0.33 & CW\\
			\quad \quad 4 & 0.36 - 0.46 & 0.31 - 0.36 & CW & 0.26 - 0.33 & 0.13 - 0.20 & ACW\\
			\quad \quad 5 & 0.66 - 0.72 & 0.23 - 0.30 & CW & 0.46 - 0.59 & 0.22 - 0.38 & ACW\\
			\quad \quad 6 & 0.60 - 0.69 & 0.15 - 0.21 & CW & 0.22 - 0.28 & 0.21 - 0.31 & CW\\
			\quad \quad 7 & 0.67 - 0.70 & 0.15 - 0.23 & CW & 0.41 - 0.49 & 0.16 - 0.22 & CW\\
			\quad \quad 8 & 0.41 - 0.62 & 0.19 - 0.30 & CW & 0.21 - 0.29 & 0.23 - 0.28 & CW\\
			\quad \quad 9 & 0.53 - 0.66 & 0.16 - 0.21 & CW & 0.22 - 0.29 & 0.18 - 0.24 & CW\\
		\end{tabular}
		\normalsize
		\caption{Radius $b_i$, distance from the arena centre $c_i$, and orientation data from the evolution of each of the two mills ($i \in \{ 1,2 \}$) in nine experimentally observed binary circular mill systems. CW is clockwise while ACW is anticlockwise.}
		\label{table:2}
	\end{center}
\end{table}

\section{Circular Milling Mathematical Model}
\subsection{$\frac{\partial u_i}{\partial x_j}$ In Bipolar Coordinates} \label{hopefullycorrect}
\begin{equation}
	\setlength{\arraycolsep}{5pt}
	\renewcommand{\arraystretch}{1.3}
	\frac{\partial u_i}{\partial x_j} = \left[ 
	\begin{array}{ccc}
	\frac{1}{h}\frac{\partial u_{\eta}}{\partial \eta} -  \frac{\sin{\xi}}{a}u_{\xi}  & \frac{1}{h}\frac{\partial u_{\eta}}{\partial \xi} +  \frac{\sinh{\eta}}{a}u_{\xi} & \frac{\partial u_{\eta}}{\partial z} \\  \frac{1}{h}\frac{\partial u_{\xi}}{\partial \eta} +  \frac{\sin{\xi}}{a}u_{\eta} & \frac{1}{h}\frac{\partial u_{\xi}}{\partial \xi} -  \frac{\sinh{\eta}}{a}u_{\eta} & \frac{\partial u_{\xi}}{\partial z} \\ \frac{\partial u_z}{\partial \eta} & \frac{\partial u_z}{\partial \xi} & \frac{\partial u_z}{\partial z}  
	\end{array} \right]. 
	\end{equation}
\subsection{$g_i$ and $I_i$} \label{algbera}
	\begin{eqnarray}
	g_0 &=& -28\, F\, \cosh\!\left(\mathrm{\eta}\right) + 28\, E\, \cosh\!\left(\mathrm{\eta}\right) + 16\, A\, \sinh\!\left(\mathrm{\eta}\right) + 8\, B\, \sinh\!\left(\mathrm{\eta}\right) - 28\, G\, \sinh\!\left(\mathrm{\eta}\right) \nonumber \\
	&+& 12\, E\, \cosh\!\left(3\, \mathrm{\eta}\right) - 32\, F\, \cosh\!\left(3\, \mathrm{\eta}\right) + 8\, A\, \sinh\!\left(3\, \mathrm{\eta}\right) - 32\, G\, \sinh\!\left(3\, \mathrm{\eta}\right) \nonumber \\
	&+& 8\, E\, \mathrm{\eta}\, \sinh\!\left(\mathrm{\eta}\right). \\
	g_1 &=& 2\, C\, - 24\, E\, + 15\, F\, - 2\, C\, \cosh\!\left(2\, \mathrm{\eta}\right)\, - 34\, E\, \cosh\!\left(2\, \mathrm{\eta}\right)\, - 2\, E\, \cosh\!\left(4\, \mathrm{\eta}\right)\, \nonumber \\
	&+& 70\, F\, \cosh\!\left(2\, \mathrm{\eta}\right)\, + 19\, F\, \cosh\!\left(4\, \mathrm{\eta}\right)\, - 26\, A\, \sinh\!\left(2\, \mathrm{\eta}\right)\, - 2\, A\, \sinh\!\left(4\, \mathrm{\eta}\right)\, \nonumber \\
	&-& 2\, B\, \sinh\!\left(2\, \mathrm{\eta}\right)\, + 70\, G\, \sinh\!\left(2\, \mathrm{\eta}\right)\, + 19\, G\, \sinh\!\left(4\, \mathrm{\eta}\right)\, - 2\, E\, \mathrm{\eta}\, \sinh\!\left(2\, \mathrm{\eta}\right). \, \\
	g_2 &=& - 28\, F\, \cosh\!\left(3\, \mathrm{\eta}\right)\, - 4\, F\, \cosh\!\left(5\, \mathrm{\eta}\right)\, + 4\, A\, \sinh\!\left(3\, \mathrm{\eta}\right)\, - 28\, G\, \sinh\!\left(3\, \mathrm{\eta}\right)\, \nonumber \\
	&-& 4\, G\, \sinh\!\left(5\, \mathrm{\eta}\right)\, + 24\, E\, \cosh\!\left(\mathrm{\eta}\right) - 32\, F\, \cosh\!\left(\mathrm{\eta}\right) + 12\, A\, \sinh\!\left(\mathrm{\eta}\right) \nonumber \\
	&-& 8\, B\,  \sinh\!\left(\mathrm{\eta}\right) - 24\, G\, \sinh\!\left(\mathrm{\eta}\right) - 8\, E\, \mathrm{\eta}\, \sinh\!\left(\mathrm{\eta}\right). \\
	g_3 &=& - 2\, C\, - 6\, E\, + 9\, F\, + 2\, C\, \cosh\!\left(2\, \mathrm{\eta}\right)\, + 2\, E\, \cosh\!\left(2\, \mathrm{\eta}\right)\, + 10\, F\, \cosh\!\left(2\, \mathrm{\eta}\right)\, \nonumber \\
	&+& 5\, F\, \cosh\!\left(4\, \mathrm{\eta}\right)\, - 2\, A\, \sinh\!\left(2\, \mathrm{\eta}\right)\, + 2\, B\, \sinh\!\left(2\, \mathrm{\eta}\right)\, + 10\, G\, \sinh\!\left(2\, \mathrm{\eta}\right)\, \nonumber \\
	&+& 5\, G\, \sinh\!\left(4\, \mathrm{\eta}\right)\, + 2\, E\, \mathrm{\eta}\, \sinh\!\left(2\, \mathrm{\eta}\right). \, \\
	g_4 &=& - 4\, F\, \cosh\!\left(\mathrm{\eta}\right) - 4\, G\, \sinh\!\left(\mathrm{\eta}\right). \\
	I_0 &=& \frac{1 + 2 \cosh^2(\eta)}{\sinh^5(\eta)}, \quad I_1 = \frac{3 \cosh(\eta)}{\sinh^5(\eta)}, \quad I_2 = \frac{3}{\sinh^5(\eta)}. \, \\
	I_3 &=& \frac{\cosh(\eta)}{\sinh^5(\eta)}(8\cosh^4(\eta) - 20\cosh^2(\eta) + 15) - 8. \\
	I_4 &=& \frac{3}{\sinh^5(\eta)}(16 \cosh^6(\eta) - 40\cosh^4(\eta) + 30\cosh^2(\eta) - 5) - 48\cosh(\eta).
	\end{eqnarray}
	

\bibliographystyle{jfm}
\bibliography{worms}

\end{document}